\documentclass[10pt,twocolumn,twoside,openbib]{article}

\usepackage{amssymb,amsmath,mathrsfs}
\usepackage{multirow}
\usepackage{graphicx}
\usepackage{float}
\usepackage{color}

\usepackage[footnotesize]{caption}
\newcommand{\icar}[1]{}
\newcommand{\prep}[1]{#1}

\def\llabel#1{\label{#1}}

\def\pdf#1{#1}

\prep{\setlength{\textwidth}{ 18cm}}
\prep{\setlength{\textheight}{  22.cm}}

\icar{\setlength{\textwidth}{ 16cm}}
\icar{\setlength{\textheight}{ 20.cm}}
\icar{\setlength{\oddsidemargin}{  0pt}}
\icar{\setlength{\evensidemargin}{  0pt}}
\prep{\setlength{\oddsidemargin}{  -1cm}}
\prep{\setlength{\evensidemargin}{ -1cm}}
\prep{\setlength{\topmargin}{  -2cm}}
\def\etal{{et al.}}
 
\newcommand{\blanc}[1]{}
\newcommand{\arcsec}{{}^{\prime\prime}}

\def\tM{{\tilde{M}}}
\def\uM{{\underline{M}}}

\def\tr{{\bf\tilde r}}

\def\e{{\bf \rm e}}

\def\bL{{\bf L}}
\def\bl{{\boldsymbol{\ell}}}
\def\rbl{{\bf r}}

\def\w{{\bf w}}                                                             
                                                             
\def\er{{\bf u}}
\def\N{{\bf N}}
                             
\def\norm#1{\left\Vert#1\right\Vert}

\newcommand{\mr}{\mathfrak{r}}
\newcommand{\ms}{\mathfrak{s}}

\newcommand{\ka}{k_3}
\newcommand{\kb}{k_5}

\newcommand{\cP}{{\cal P}}
\newcommand{\cS}{{\cal S}}
\newcommand{\cQ}{{\cal Q}}
\newcommand{\cC}{{\mathscr{C}}}

\newcommand{\fA}{{{A'}}}
\newcommand{\fC}{{\mathcal{C}}}
\newcommand{\fT}{{\mathcal{D}}}
\newcommand{\fCa}{{\mathcal{C}_1^\prime}}
\newcommand{\fCb}{{\mathcal{C}_2^\prime}}

\newcommand{\fTa}{{\mathcal{D}_1^\prime}}
\newcommand{\fTb}{{\mathcal{D}_2^\prime}}

\newcommand{\mz}{\mathfrak{z}}

\newcommand{\cX}{{\cal X}}
\newcommand{\cY}{{\cal Y}}

\newcommand{\cW}{{\cal W}}
\newcommand{\cR}{{\cal R}}
\newcommand{\cL}{{\cal L}}
\newcommand{\cA}{{\cal A}}
\newcommand{\LA}{\left\langle}
\newcommand{\RA}{\right\rangle}

\newcommand{\BR}{\mathbb{R}}
\newcommand{\cH}{{\cal H}}
\newcommand{\Ham}{\cH}

\newcommand{\He}{H_{{E}}}

\newcommand{\Hi}{H_{{I}}}
\newcommand{\Ht}{H_T}
\newcommand{\Hs}{H_s}
\newcommand{\tHs}{\underline{H}_s}

\newcommand{\s}{\boldsymbol{s}}
\newcommand{\cG}{\mathcal{G}}
\newcommand{\cB}{\mathcal{B}}
\newcommand{\bM}{\boldsymbol{M}}
\newcommand{\G}{\boldsymbol{G}}
\newcommand{\U}{\bf U}

\newcommand{\bW}{\boldsymbol{W}}

\newcommand{\bi}{\boldsymbol{i}}
\newcommand{\bj}{\boldsymbol{j}}
\newcommand{\bk}{\boldsymbol{k}}

\newcommand{\n}{\boldsymbol{n}}
\newcommand{\bI}{\boldsymbol{I}}
\newcommand{\bJ}{\boldsymbol{J}}
\newcommand{\bK}{\boldsymbol{K}}
\newcommand{\bG}{\boldsymbol{G}}

\newcommand{\Id}{Id}
\newcommand{\In}{\mathcal{I}}

\newcommand{\trans}[1]{{}^t{#1}}

\newcommand{\dop}[2]{{#1}\!\cdot{#2}}
\newcommand{\grad}[1]{\boldsymbol{\nabla}_{\!#1}}

\newcommand{\fa}{\mathfrak{a}}
\newcommand{\fb}{\mathfrak{b}}
\newcommand{\fc}{\mathfrak{c}}
\newcommand{\fd}{\mathfrak{d}}
\newcommand{\fe}{\mathfrak{e}}
\newcommand{\ff}{\mathfrak{f}}
\newcommand{\fg}{\mathfrak{g}}
\newcommand{\fh}{\mathfrak{h}}
\newcommand{\fu}{\mathfrak{u}}

\newcommand{\tfa}{\underline{\mathfrak{a}}}
\newcommand{\tfb}{\underline{\mathfrak{b}}}
\newcommand{\tbT}{\underline{\bf T}}
\newcommand{\tbP}{\underline{\bf P}}

\newcommand{\B}{\beta}

\newcommand{\A}{\gamma}

\newcommand{\const}{{Cte}}

\newcommand{\rra}{(\dop{\er}{\rbl_1})}
\newcommand{\rrb}{(\dop{\er}{\rbl_2})}
\newcommand{\rarb}{(\dop{\rbl_1}{\rbl_2})}

\def\crm{\cr\noalign{\medskip}}

\def\m@th{\mathsurround=0pt}
\def\EQM#1{\vcenter{\normalbaselines\m@th
    \ialign{${\displaystyle ##}$\hfil&&\ ${\displaystyle ##}$\hfil\crcr
    \mathstrut\crcr\noalign{\kern-\baselineskip}
    \noalign{\smallskip}
    #1\crcr\mathstrut\crcr\noalign{\kern-\baselineskip}}}}
\newcommand{\Frac}[2]{{{\displaystyle\strut#1}\over{\displaystyle\strut#2}}}
\newcommand{\be}{\begin{equation}}
\newcommand{\ee}{\end{equation}}

\def\Dron#1#2{\frac{\partial#1}{\partial#2}}
\def\bhs{\hspace{\stretch{1}}}
\def\ds{\displaystyle}

\newcommand{\bpm}{\begin{pmatrix}}
\newcommand{\epm}{\end{pmatrix}}

\newtheorem{prop}{Proposition}
\newenvironment{proof}{{\it Proof.}}{}
\newenvironment{consequence}{{\it Consequence.}}{}

\prep{}
\prep{}
\icar{}
\icar{}

\newcommand\figa{
\begin{figure}[t] 
\begin{center}
 \includegraphics[width=8cm]{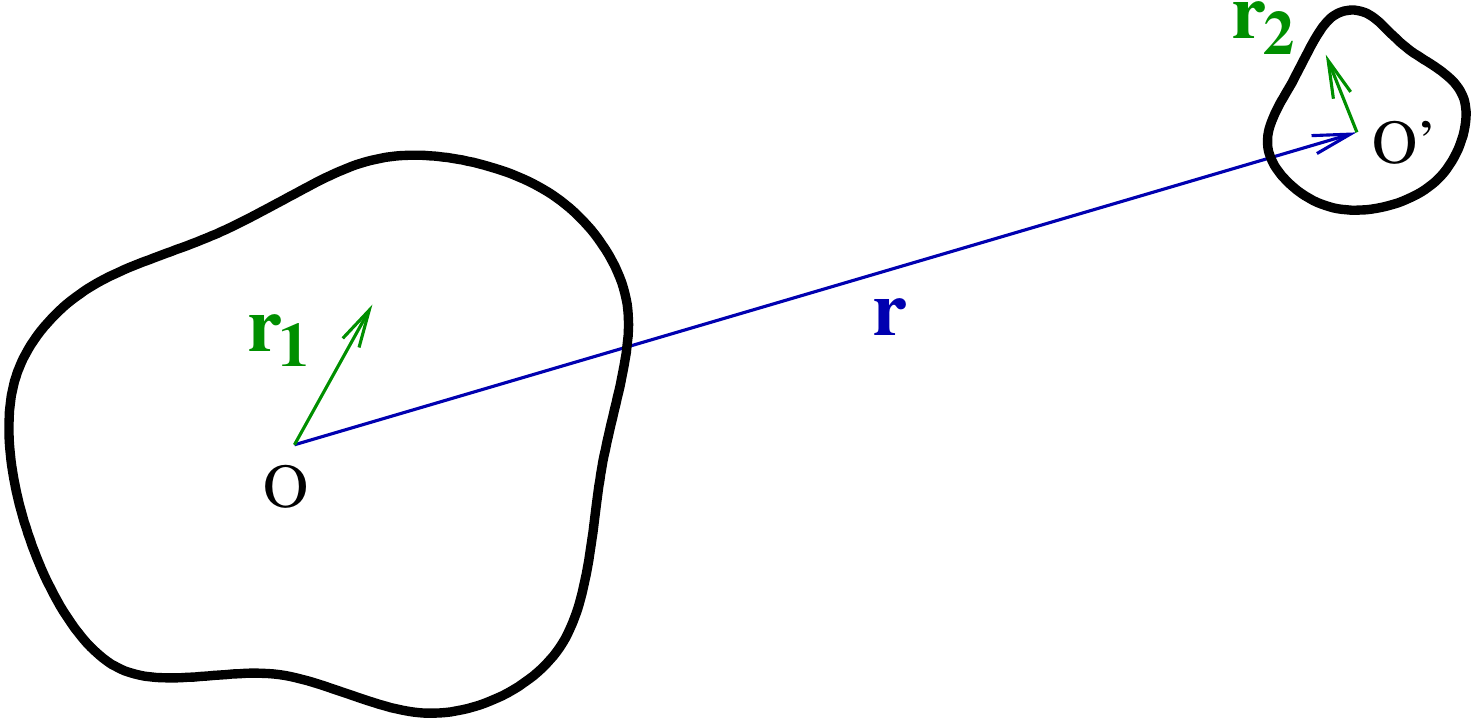} 
  \caption{Coordinates definition.}
  \llabel{Figa}
\end{center}
\end{figure}
}

\newcommand\figb{
\begin{figure}[t]
\begin{center}
\pdf{\includegraphics[width=8cm,viewport=0 150 615 615]{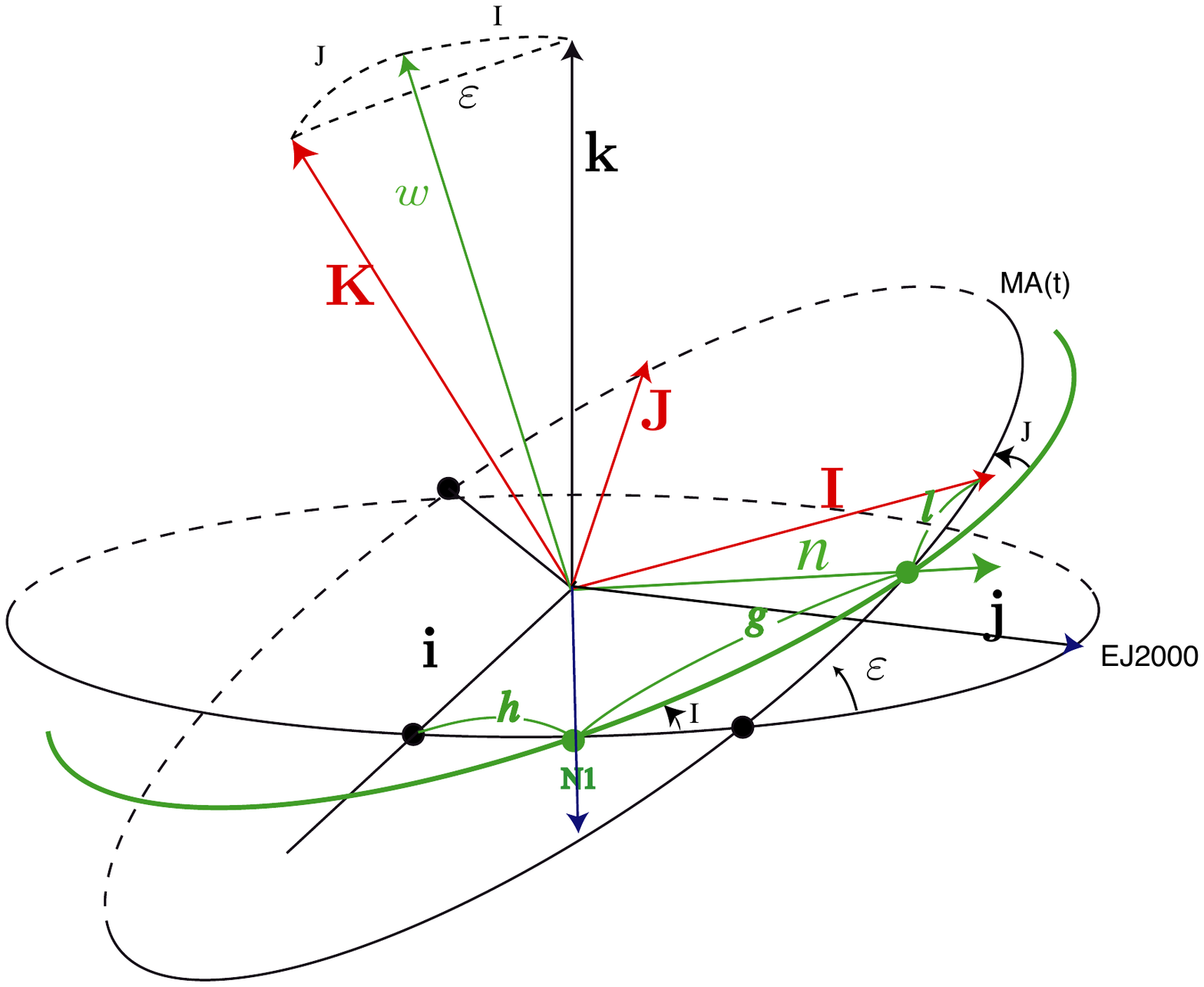}}
  \caption{Definition of Andoyer's coordinates. 
  $(\bi,\bj,\bk)$ is a fixed reference frame, 
  and $(\bI,\bJ,\bK)$ the reference frame of the principal 
  axis of inertia of the solid body. The Andoyer 
  action variables are $(G,H=\bG\cdot\bk,L=\bG\cdot\bK)$  with the associated angles 
  $(g,h,l)$ (Andoyer, 1923).}
  \llabel{Figb}
\end{center}
\end{figure}
}

\newcommand\figc{
\begin{figure}[t] 
\begin{center}
\pdf{\includegraphics[width=8cm]{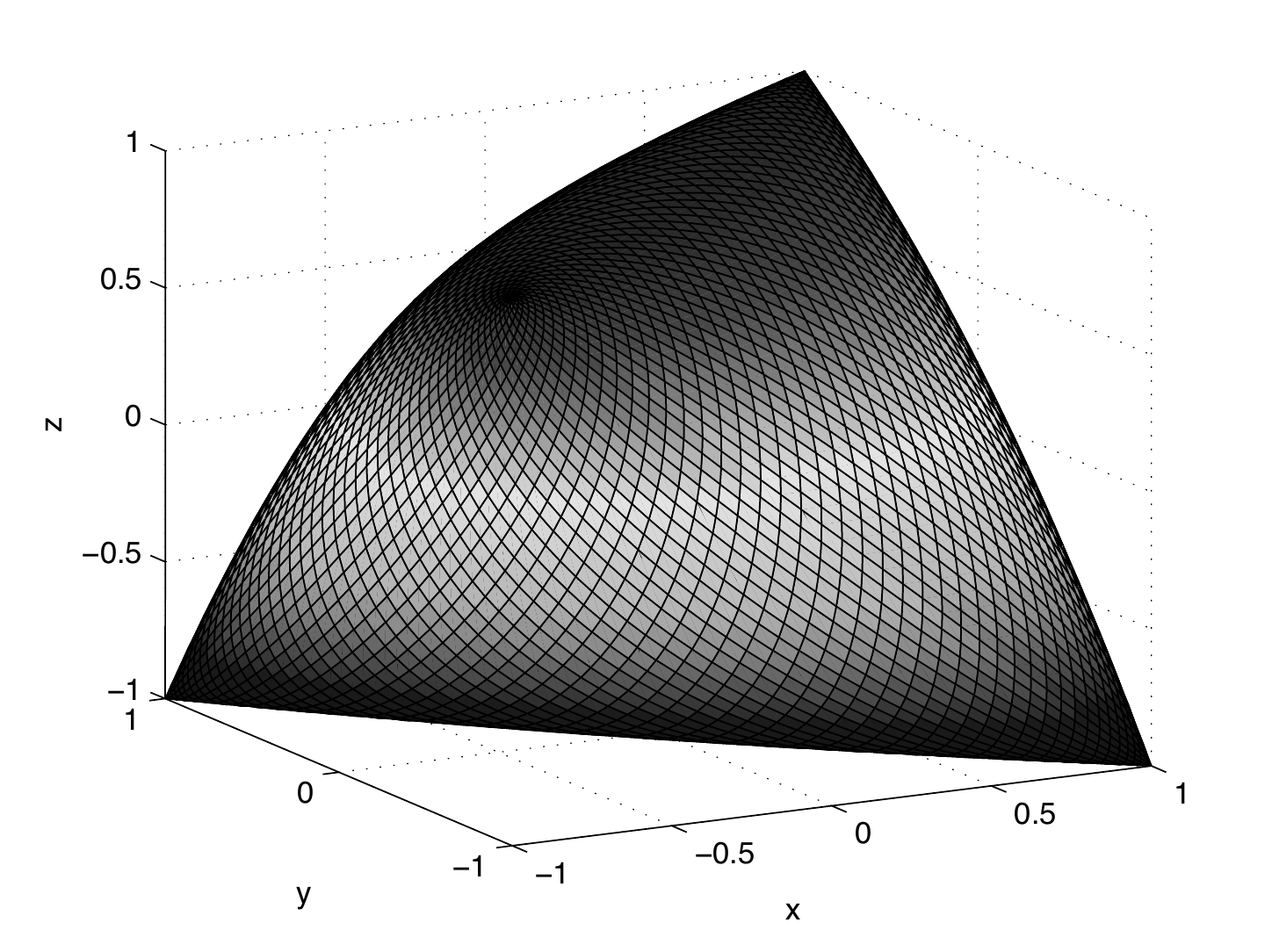}} 
  \caption{The surface $v^2(x,y,z)=0$. As $v^2 \geq 0$, the allowed 
  space is the interior of this Cassini berlingot shaped  volume.}
  \llabel{Figc}
\end{center}
\end{figure}
}

\newcommand\figca{
\begin{figure}[t]
\begin{center}
\includegraphics[height=8cm]{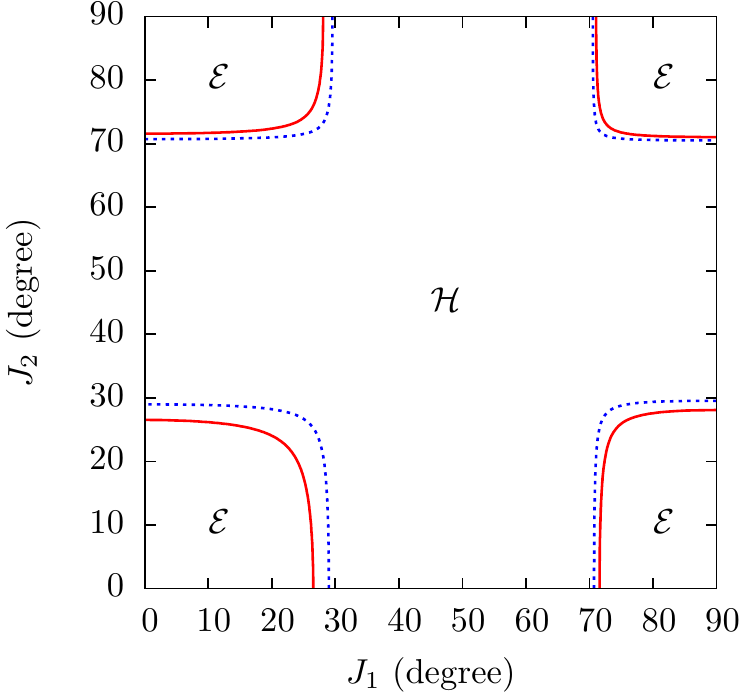}
\caption{Shape of the surface $\cal Q'$ as a function of the
angles $J_1$ and $J_2$ between the angular momentum and the axis of
maximum inertia of each bodies. $\cal E$ and $\cal H$ stand for ellipsoid and 
hyperboloid respectively. The solid curve in red delineates the $\cal E$
and the $\cal H$ zones for $\eta=25/9$. The dashed blue curve
corresponds to $\eta=25/4$. See text for details.}
\llabel{Figca}
\end{center}
\end{figure}
}

\newcommand\figd{
\begin{figure}[t] 
\begin{center}
\pdf{\includegraphics[width=5cm]{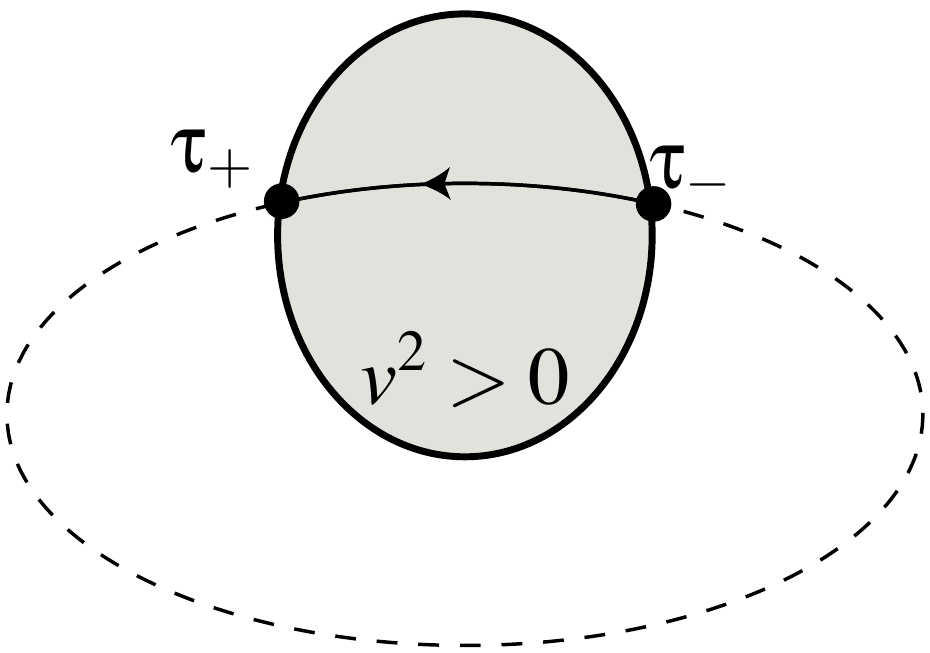}} 
  \caption{The shaded area corresponds to the region where $v^2>0$,
inside the Cassini berlingot $\cB$. The orbit 
intersects the Cassini berlingot $\cB$ in $t=\tau_+$ and $t=\tau_-$.}
  \llabel{Figd}
\end{center}
\end{figure}
}

\newcommand\fige{
\begin{figure*}[t]
\begin{center}
\includegraphics[height=8cm]{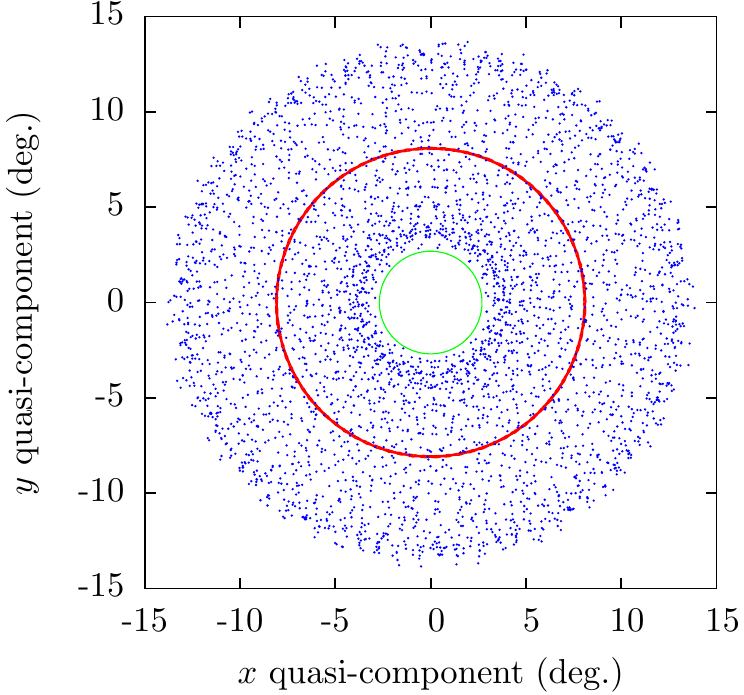}\hspace{\stretch{1}}\includegraphics[height=8cm]{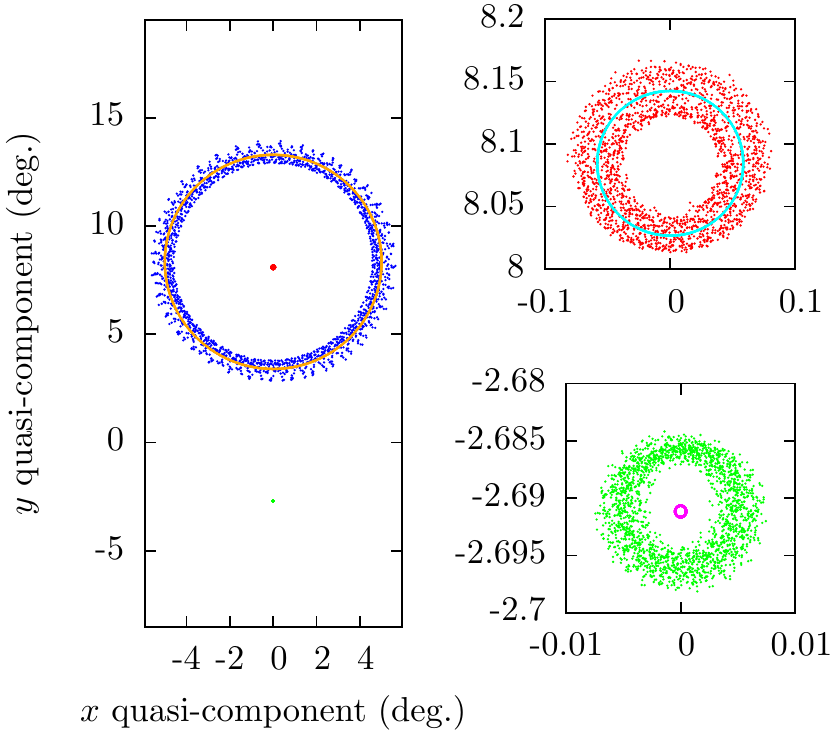}
\icar{\renewcommand{\baselinestretch}{1.1}}
\caption{Quasi-projection of the poles $\w$ (red), $\w_1$ (green), $\w_2$ (blue) 
on the plane perpendicular to the total angular momentum $\bW_0$, in a
fixed reference frame (left panel) and in a frame rotating with the
precession period (right panels). The two little figures on the right are
zooms on the nutation motion of the orbit (top) and of the primary axis
(bottom). The initial conditions and parameters
are those of the system $I$. The vectors $\w$, $\w_1$ and $\w_2$ have
been integrated with the full Hamiltonian. In the right panels, the output
of the averaged Hamiltonian has been superposed: $\w$ in cyan, $\w_1$ in pink 
and $\w_2$ in orange.
}
\icar{\renewcommand{\baselinestretch}{1.525}}
\llabel{Fige}
\end{center}
\end{figure*}
}

\newcommand\figf{
\begin{figure}[t]
\begin{center}
\includegraphics[height=8cm]{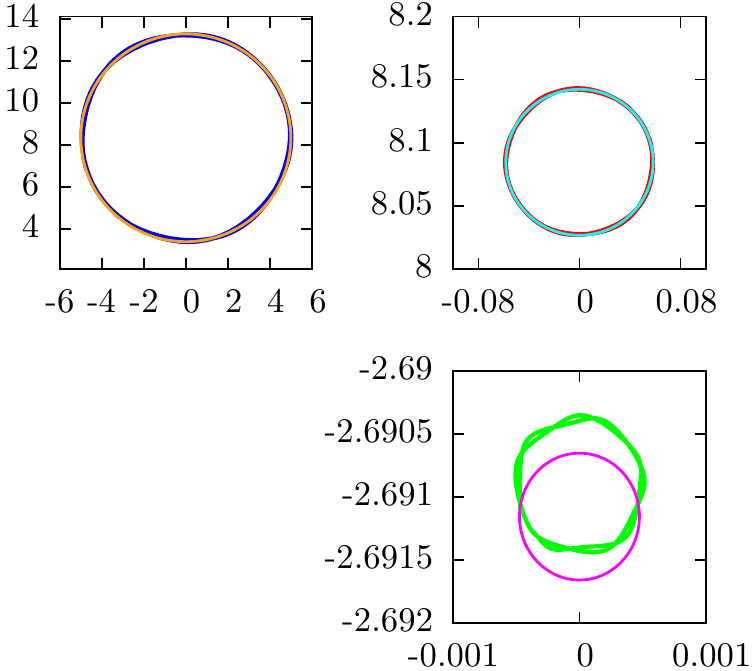}
\caption{Same as the right panel of Fig.~\ref{Fige}. The output of
the full Hamiltonian, integrated over 20 days with an output step of 30
min, has been filtered with a low-pass filter with a
cutoff frequency equal to 4 rad/day.}
\llabel{Figf}
\end{center}
\end{figure}
}

\newcommand\figg{
\begin{figure}[t]
\begin{center}
\includegraphics[height=8cm]{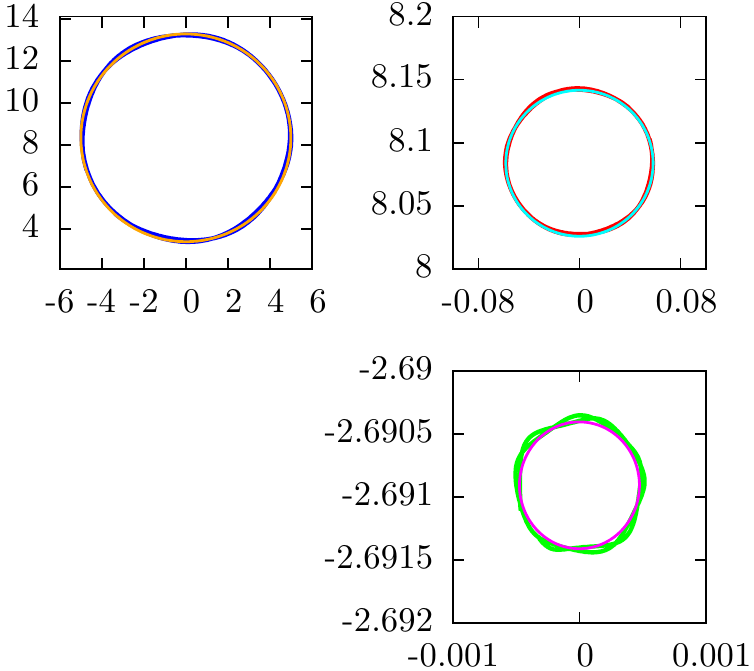}
\caption{Same as Fig.~\ref{Figf}. The initial
obliquity of the primary in the averaged Hamiltonian has been decreased
by $3.6\arcsec$.}
\llabel{Figg}
\end{center}
\end{figure}
}

\newcommand\taba{
\begin{table}[t!]
\begin{center}
  \caption{Physical and orbital parameters of a fictitious doubly asynchronous
binary system. $m$ is the mass, $A$, $B$ and $C$ are the moments of
inertia divided the mass, $w$ is the rotation rate, $h$, $I$, $g$, $J$
and $l$ are the Andoyer angles of the two solid bodies as defined in Fig.~\ref{Figb}.}
\begin{tabular}{lrr|lr}
\multicolumn{5}{c}{System $I$} \\
\hline
                             &    Primary     &    Secondary    &                 & Orbit      \\
\hline                      
$m$ ($10^{12}$kg)            &  $ 2.5     $ & $   0.15   $ & $a$ (km)        & $ 2.75 $ \\    
$A$ (km$^2$)                 &  $ 0.17    $ & $   0.0165 $ & $\lambda$ (deg) & $ 0.0  $ \\    
$B$ (km$^2$)                 &  $ 0.18    $ & $   0.017  $ & $e$             & $ 0.035$ \\    
$C$ (km$^2$)                 &  $ 0.19    $ & $   0.025  $ & $\omega$ (deg)  & $ 0.0  $ \\    
$w$ ($^\circ$/day)           &  $ 3125.34 $ & $  1500    $ & $i$ (deg)       & $ 0.0  $ \\    
$h$ (deg)                    &  $ 100.82  $ & $  -110.0  $ & $\Omega$ (deg)  & $ 0.0  $ \\    
$I$ (deg)                    &  $ 10.74   $ & $     5.0  $ &                 &          \\    
$g$ (deg)                    &  $ 112.03  $ & $  -180.0  $ &                 &          \\    
$J$ (deg)                    &  $  3.0    $ & $     5.0  $ &                 &          \\
$l$ (deg)                    &  $ 90.0    $ & $    90.0  $ &                 &          \\
\hline
\end{tabular}
  \llabel{Taba}
\end{center}
\end{table}
}

\newcommand\tabab{
\begin{table}[t!]
\begin{center}
  \caption{Physical and orbital parameters of the binary asteroids 1999
KW4 given by FS08. $m$ is the mass, $A$, $B$ and $C$ are the moments of
inertia divided the mass, $w$ is the rotation rate, $h$, $I$, $g$, $J$
and $l$ are the Andoyer angles of the two solid bodies as defined in Fig.~\ref{Figb}. }
\begin{tabular}{lrr|lr}
\multicolumn{5}{c}{System $I\!I$} \\
\hline
                             &    Primary     &    Secondary    &                 & Orbit      \\
\hline                      
$m$ ($10^{12}$kg)       &  $ 2.353     $ & $   0.135     $ & $a$ (km)        & $ 2.5405 $ \\    
$A$ (km$^2$)            &  $ 0.1648    $ & $   0.01608   $ & $\lambda$ (deg) & $ 0.0    $ \\    
$B$ (km$^2$)            &  $ 0.1726    $ & $   0.02374   $ & $e$             & $ 0.0    $ \\    
$C$ (km$^2$)            &  $ 0.1959    $ & $   0.02799   $ & $\omega$ (deg)  & $ 0.0    $ \\    
$w$ ($^\circ$/day)      &  $ 3125.34   $ & $   498.09    $ & $i$ (deg)       & $ 0.0    $ \\    
$h$ (deg)               &  $ 117.04    $ & $   0.0       $ & $\Omega$ (deg)  & $ 0.0    $ \\    
$I$ (deg)               &  $ 10.0      $ & $   0.0       $ &                 &            \\    
$g$ (deg)               &  $  0.0      $ & $   0.0       $ &                 &            \\    
$J$ (deg)               &  $  0.0      $ & $   0.0       $ &                 &            \\    
$l$ (deg)               &  $-173.93    $ & $   180.0     $ &                 &            \\    
\hline
\end{tabular}
  \llabel{Tabab}
\end{center}
\end{table}
}

\newcommand\tabba{
\begin{table*}[t!]
\icar{\renewcommand{\baselinestretch}{0.8}}
\begin{center}
 \caption{Frequency decomposition of the motion of the projections 
$\mz$, $\mz_1$ and $\mz_2$ respectively of $\w$, $\w_1$ and $\w_2$ on
the plane orthogonal to the total angular momentum $\bW_0$. The
integration was made using the full Hamiltonian with the initial
conditions of the doubly asynchronous system $I$. Only the
first 20 terms of the series $\sum A_j \exp{i(\nu_j t+\varphi_j)}$ are 
displayed for each vector. In order to simplify the
reading, hats on the angles $\omega$, $g_1$, $l_1$, $g_2$ and $l_2$ are omitted.}
{\scriptsize
\begin{tabular}{*{6}{r}}
\hline
var. & $i$ &        &  $\nu_i$  & $A_i^{(f)}$ & $\varphi_i^{(f)}$ \\
     &     &        &  $(\text{rad.yr}^{-1})$ & $(\arcsec)$  & (deg) \\ \hline 
\multirow{20}{*}{$\w$} 
&  1 &$ \Omega $& -0.0312 & 29105.09 & -169.37 \\
&  2 &$ \Omega +\nu $& -1.0100 & 209.11 & -17.70 \\
&  3 &$ \Omega +2\omega +2n $& 16.1156 & 55.87 & -10.55 \\
&  4 &$ \Omega -\nu +2\omega +2n $& 17.0943 & 12.29 & -162.22 \\
&  5 &$ \Omega +2\omega +2n -g_2 $& -23.8548 & 9.55 & -158.86 \\
&  6 &$ \Omega +2\omega +2n -g_1 $& -42.8607 & 6.70 & 57.27 \\
&  7 &$ \Omega +g_2 $& 39.9391 & 5.72 & -21.06 \\
&  8 &$ \Omega -n $& -8.0365 & 5.22 & -168.69 \\
&  9 &$ \Omega +n $& 7.9740 & 5.21 & 9.94 \\
& 10 &$ \Omega +g_1 $& 58.9451 & 4.89 & 122.82 \\
& 11 &$ \Omega +2\omega +3n $& 24.1208 & 4.00 & -11.23 \\
& 12 &$ \Omega +2\omega +2n -2l_1 -2g_1 $& -93.0245 & 3.39 & -55.00 \\
& 13 &$ \Omega +2l_1 +2g_1 $& 109.1089 & 2.90 & -124.92 \\
& 14 &$ \Omega +2\omega +n $& 8.1103 & 1.70 & 170.13 \\
& 15 &$ \Omega +2\omega +3n -g_2 $& -15.8495 & 1.47 & -159.55 \\
& 16 &$ \Omega +2\omega +2n -2l_1 -g_1 $& -34.0483 & 1.39 & 57.17 \\
& 17 &$ \Omega +\nu +n $& 6.9953 & 1.08 & 161.62 \\
& 18 &$ \Omega +\nu -n $& -9.0152 & 1.04 & -17.02 \\
& 19 &$ \Omega -\nu $& 0.9476 & 0.95 & -141.05 \\
& 20 &$ \Omega +2l_1 +g_1 $& 50.1326 & 0.94 & 122.90 \\
\hline
\multirow{20}{*}{$\w_1$} 
&  1 &$ \Omega $& -0.0312 & 9687.25 & 10.63 \\
&  2 &$ \Omega +2\omega +2n $& 16.1156 & 18.80 & 169.45 \\
&  3 &$ \Omega +2\omega +2n -g_1 $& -42.8607 & 2.26 & -122.72 \\
&  4 &$ \Omega +\nu $& -1.0100 & 1.86 & 162.30 \\
&  5 &$ \Omega -n $& -8.0365 & 1.73 & 11.31 \\
&  6 &$ \Omega +n $& 7.9740 & 1.73 & -170.06 \\
&  7 &$ \Omega +g_1 $& 58.9451 & 1.60 & -57.20 \\
&  8 &$ \Omega +2\omega +3n $& 24.1208 & 1.34 & 168.77 \\
&  9 &$ \Omega -2\omega -2n +2l_1 +2g_1 $& 92.9621 & 1.11 & 76.25 \\
& 10 &$ \Omega +2l_1 +2g_1 $& 109.1089 & 0.94 & 55.08 \\
& 11 &$ \Omega +2\omega +n $& 8.1103 & 0.58 & -9.87 \\
& 12 &$ \Omega +2\omega +2n -2l_1 -g_1 $& -34.0483 & 0.47 & -122.83 \\
& 13 &$ \Omega +2l_1 +g_1 $& 50.1326 & 0.31 & -57.10 \\
& 14 &$ \Omega +2\omega +3n -g_1 $& -34.8555 & 0.30 & -123.41 \\
& 15 &$ \Omega -2\omega -2n $& -16.1780 & 0.16 & -148.19 \\
& 16 &$ \Omega -2\omega -3n +2l_1 +2g_1 $& 84.9569 & 0.13 & 76.94 \\
& 17 &$ \Omega -\nu +2\omega +2n $& 17.0943 & 0.13 & 17.78 \\
& 18 &$ \Omega -n +g_1 $& 50.9398 & 0.08 & -56.52 \\
& 19 &$ \Omega +2\omega +4n $& 32.1260 & 0.07 & 168.08 \\
& 20 &$ \Omega +2\omega +3n -2l_1 -g_1 $& -26.0430 & 0.07 & -123.51 \\
\hline
\multirow{20}{*}{$\w_2$} 
&  1 &$ \Omega $& -0.0312 & 30079.18 & -169.37 \\
&  2 &$ \Omega +\nu $& -1.0100 & 17781.88 & 162.30 \\
&  3 &$ \Omega -\nu +2\omega +2n $& 17.0943 & 1040.78 & 17.78 \\
&  4 &$ \Omega +2\omega +2n -g_2 $& -23.8548 & 829.13 & 21.13 \\
&  5 &$ \Omega +g_2 $& 39.9391 & 495.19 & 158.94 \\
&  6 &$ \Omega +2\omega +3n -g_2 $& -15.8495 & 126.95 & 20.45 \\
&  7 &$ \Omega -\nu $& 0.9476 & 96.54 & 38.96 \\
&  8 &$ \Omega +\nu +n $& 6.9953 & 90.98 & -18.39 \\
&  9 &$ \Omega +\nu -n $& -9.0152 & 87.74 & 162.98 \\
& 10 &$ \Omega -\nu +2\omega +3n $& 25.0996 & 74.38 & 17.10 \\
& 11 &$ \Omega -\nu +2\omega +n $& 9.0891 & 50.41 & -161.54 \\
& 12 &$ \Omega +2\omega +2n $& 16.1156 & 46.84 & -10.55 \\
& 13 &$ \Omega -n +g_2 $& 31.9339 & 25.33 & 159.63 \\
& 14 &$ \Omega +2\omega +2n -2l_2 -g_2 $& 3.9537 & 23.66 & -161.23 \\
& 15 &$ \Omega +n +g_2 $& 47.9444 & 21.72 & 158.26 \\
& 16 &$ \Omega +2\omega +4n -g_2 $& -7.8443 & 19.43 & 19.77 \\
& 17 &$ \Omega +2n +2l_1 -2l_2 $& 34.9754 & 14.90 & -57.53 \\
& 18 &$ \Omega -\nu +2l_2 +2g_2 $& 53.0797 & 10.87 & -22.05 \\
& 19 &$ \Omega +2l_2 +g_2 $& 12.1306 & 8.74 & 161.31 \\
& 20 &$ \Omega +2\omega +5n -g_2 $& 0.1610 & 7.16 & -160.92 \\
\hline
  \end{tabular}
}
  \llabel{Tabba}
\icar{\renewcommand{\baselinestretch}{1.525}}
\end{center}
\end{table*}
}

\newcommand\tabbb{
\begin{table*}[t!]
\icar{\renewcommand{\baselinestretch}{0.8}}
\begin{center}
  \caption{Same as table~\ref{Tabba} for the system $I\!I$.}
{\scriptsize
\begin{tabular}{*{6}{r}}
\hline
var. & $i$ &        &  $\nu_i$  & $A_i^{(f)}$ & $\varphi_i^{(f)}$ \\
    &     &        &  $(\text{rad.yr}^{-1})$ & $(\arcsec)$  & (deg) \\ \hline 
\multirow{20}{*}{$\w$} 
&  1 &$ \Omega $& -0.0713 & 27979.553 & -153.15 \\
&  2 &$ \Omega +2\omega +2n $& 17.8490 & 111.214 & -26.81 \\
&  3 &$ \Omega +\nu $& -4.8201 & 4.421 & -160.28 \\
&  4 &$ \Omega +2\omega +2n -2l_1 -2g_1 $& -91.3906 & 3.320 & -39.33 \\
&  5 &$ \Omega -n $& -9.1216 & 2.847 & 25.69 \\
&  6 &$ \Omega +2l_1 +2g_1 $& 109.1682 & 2.779 & -140.63 \\
&  7 &$ \Omega +n $& 8.9790 & 2.678 & -151.99 \\
&  8 &$ \Omega +2\omega +3n $& 26.8993 & 2.201 & 154.37 \\
&  9 &$ \Omega +\psi_2 $& 7.5201 & 1.295 & -152.20 \\
& 10 &$ \Omega -\psi_2 $& -7.6627 & 1.117 & 25.96 \\
& 11 &$ \Omega +2\omega +n $& 8.7986 & 1.066 & -27.97 \\
& 12 &$ \Omega -\nu +2\omega +2n $& 22.5978 & 0.942 & -19.67 \\
& 13 &$ \Omega +2\omega+2n+\psi_2 $& 25.4404 & 0.828 & 155.16 \\
& 14 &$ \Omega -2\omega -2n $& -17.9916 & 0.490 & -99.49 \\
& 15 &$ \Omega +\omega +n -\theta_2 $& 4.7413 & 0.376 & -163.33 \\
& 16 &$ \Omega +2\omega +2n -\psi_2 $& 10.2575 & 0.280 & -27.75 \\
& 17 &$ \Omega +\omega -\theta_2 $& -4.3090 & 0.237 & -164.49 \\
& 18 &$ \Omega -\nu +2\omega -n $& -4.5532 & 0.170 & 156.75 \\
& 19 &$ \Omega -\nu +2\omega $& 4.4971 & 0.161 & 157.74 \\
& 20 &$ \Omega +\nu +n $& 4.2302 & 0.143 & 20.77 \\
\hline
\multirow{20}{*}{$\w_1$} 
&  1 &$ \Omega $& -0.0713 & 8008.982 & 26.85 \\
&  2 &$ \Omega +2\omega +2n $& 17.8490 & 32.172 & 153.19 \\
&  3 &$ \Omega -2\omega -2n +2l_1 +2g_1 $& 91.2480 & 0.939 & 93.03 \\
&  4 &$ \Omega -n $& -9.1216 & 0.796 & -154.31 \\
&  5 &$ \Omega +n $& 8.9790 & 0.794 & 28.01 \\
&  6 &$ \Omega +2l_1 +2g_1 $& 109.1682 & 0.780 & 39.37 \\
&  7 &$ \Omega +2\omega +3n $& 26.8993 & 0.636 & -25.64 \\
&  8 &$ \Omega -\psi_2 $& -7.6627 & 0.325 & -154.10 \\
&  9 &$ \Omega +\psi_2 $& 7.5201 & 0.324 & 27.80 \\
& 10 &$ \Omega +2\omega +n $& 8.7986 & 0.301 & 152.03 \\
& 11 &$ \Omega -2\omega -2n $& -17.9916 & 0.243 & -99.49 \\
& 12 &$ \Omega +2\omega +2n +\psi_2 $& 25.4404 & 0.232 & -25.86 \\
& 13 &$ \Omega +2\omega +2n -\psi_2 $& 10.2575 & 0.093 & 152.25 \\
& 14 &$ \Omega -2\omega -3n +2l_1 +2g_1 $& 82.1976 & 0.031 & -88.11 \\
& 15 &$ \Omega +\nu $& -4.8201 & 0.030 & 19.72 \\
& 16 &$ \Omega +4\omega+4n-2l_1-2g_1 $& -73.4703 & 0.029 & 87.00 \\
& 17 &$ \Omega +2\omega $& -0.2517 & 0.024 & -29.13 \\
& 18 &$ \Omega +2\omega +2n -2l_1 -2g_1 $& -91.3906 & 0.023 & 140.67 \\
& 19 &$ \Omega +n -\psi_2 $& 1.3876 & 0.020 & -152.96 \\
& 20 &$ \Omega -n +\psi_2 $& -1.5302 & 0.020 & 26.63 \\
\hline
\multirow{20}{*}{$\w_2$} 
&  1 &$ \Omega $& -0.0713 & 28848.685 & -153.15 \\
&  2 &$ \Omega +\nu $& -4.8201 & 924.226 & 19.72 \\
&  3 &$ \Omega -\nu +2\omega +2n $& 22.5978 & 196.275 & 160.33 \\
&  4 &$ \Omega +2\omega +2n $& 17.8490 & 81.967 & -26.85 \\
&  5 &$ \Omega +\omega +n -\theta_2 $& 4.7413 & 77.510 & 16.67 \\
&  6 &$ \Omega +\nu +n $& 4.2302 & 55.301 & -159.15 \\
&  7 &$ \Omega +\omega -\theta_2 $& -4.3090 & 47.469 & 15.51 \\
&  8 &$ \Omega -\nu +2\omega -n $& -4.5532 & 34.386 & -23.16 \\
&  9 &$ \Omega -\nu +2\omega $& 4.4971 & 33.518 & -22.15 \\
& 10 &$ \Omega +\nu -\psi_2 $& -12.4116 & 30.128 & 18.77 \\
& 11 &$ \Omega +\omega +n +\theta_2 $& 13.0363 & 29.773 & -16.63 \\
& 12 &$ \Omega -\psi_2 $& -7.6627 & 29.524 & -151.52 \\
& 13 &$ \Omega +\omega +\theta_2 $& 3.9860 & 26.699 & 162.21 \\
& 14 &$ \Omega +\nu -n $& -13.8705 & 26.395 & -161.45 \\
& 15 &$ \Omega +\omega -n +\theta_2 $& -5.0643 & 19.869 & -18.95 \\
& 16 &$ \Omega +\nu +\psi_2 $& 2.7713 & 19.196 & 20.64 \\
& 17 &$ \Omega +\omega +n -\psi_2 +\theta_2 $& 5.4449 & 18.432 & 162.43 \\
& 18 &$ \Omega +\nu +2n $& 13.2805 & 13.450 & 22.33 \\
& 19 &$ \Omega -\nu +2\omega +2n -\psi_2 $& 15.0063 & 12.645 & 159.39 \\
& 20 &$ \Omega +\omega +n -\psi_2 -\theta_2 $& -2.8501 & 12.132 & -164.27 \\
\hline
  \end{tabular}
}
  \llabel{Tabbb}
\end{center}
\icar{\renewcommand{\baselinestretch}{1.525}}
\end{table*}
}

\newcommand\tabc{
\begin{table*}[t!]
\begin{center}
  \caption{Frequency analysis of the doubly asynchronous system $I$.
Columns 3 to 5 correspond to the frequency analysis performed on the
numerical integration of the averaged hamiltonian (\ref{eq.Hs}),
superscript $(a)$. Columns 6 to 8 contain the secular terms of the
frequency decompositions computed on the output of the full integration,
superscript $(f)$. Columns 9 to 11 are the results of the analytical
approximations (\ref{eq.omeganu} and \ref{eq.ini}), superscript $(c)$.} 
\begin{tabular}{l*{10}{r}}
\hline
var. &
& $i$  &  $A_i^{(a)}$ & $\varphi_i^{(a)}$  & 
$i$ &     $A_i^{(f)}$ & $\varphi_i^{(f)}$  &
$i$ &     $A_i^{(c)}$ & $\varphi_i^{(c)}$  \\
&   &  &  $(\arcsec)$ & (deg)              & 
    &     $(\arcsec)$ & (deg)              & 
    &     $(\arcsec)$ & (deg)              \\ \hline 
\multirow{4}{*}{$\w$}
&$\Omega         $&  1 & 29104.12  & -169.37 &  1 & 29105.09  & -169.37 & 1 & 29108.16  &  -169.37 \\
&$\Omega +\nu    $&  2 & 209.34    &  -17.37 &  2 &   209.11  &  -17.70 & 2 &   212.43  &   -17.10 \\
&$\Omega -\nu    $&  3 & 0.95      & -141.38 & 19 &     0.95  & -141.05 &   &           &          \\
&$\Omega +2\nu   $&  4 & 0.04      &  134.64 & 57 &     0.04  &  134.09 &   &           &          \\
\hline 
\multirow{3}{*}{$\w_1$}
&$\Omega         $&  1 & 9688.12   &   10.63 &  1 &  9687.25  &  10.63  & 1 &  9688.16  &    10.63 \\
&$\Omega +\nu    $&  2 & 1.76      &  162.63 &  4 &     1.86  &  162.30 & 2 &     1.85  &   162.90 \\
&$\Omega -\nu    $&  3 & 0.05      & -141.38 & 23 &     0.05  & -141.02 &   &           &          \\
\hline 
\multirow{5}{*}{$\w_2$}
&$\Omega         $&  1 & 30020.39  & -169.37 &  1 & 30079.18  & -169.37 & 1 & 29676.07  &  -169.37 \\
&$\Omega +\nu    $&  2 & 17889.69  &  162.63 &  2 & 17781.88  &  162.30 & 2 & 18137.11  &   162.90 \\
&$\Omega -\nu    $&  3 & 96.21     &   38.63 &  7 &    96.54  &   38.96 &   &           &          \\
&$\Omega +2\nu   $&  4 & 3.29      &  -45.36 & 37 &     2.95  &  -46.02 &   &           &          \\
&$\Omega -2\nu   $&  5 & 0.02      & -113.38 &    &           &         &   &           &          \\
\hline
  \end{tabular}
  \llabel{Tabc}
\end{center}
\end{table*}
}

\newcommand\tabd{
\begin{table*}[t!]
\begin{center}
  \caption{Same as table~\ref{Tabc} for the system $I\!I$ corresponding
to the 1999 KW4 binary asteroids.} 
\begin{tabular}{l*{10}{r}}
\hline
var. &
& $i$  &  $A_i^{(a)}$ & $\varphi_i^{(a)}$  & 
$i$ &     $A_i^{(f)}$ & $\varphi_i^{(f)}$  &
$i$ &     $A_i^{(c)}$ & $\varphi_i^{(c)}$  \\
&   &  &  $(\arcsec)$ & (deg)              & 
    &     $(\arcsec)$ & (deg)              & 
    &     $(\arcsec)$ & (deg)              \\ \hline 
\multirow{3}{*}{$\w$}
&$\Omega         $&  1 & 27916.13  & -152.96 &  1 & 27979.55  & -153.15 & 1 & 27916.04  &  -152.96 \\
&$\Omega +\nu    $&  2 & 4.65      & -152.96 &  3 &     4.42  & -160.28 & 2 &     4.72  &  -152.96 \\
&$\Omega -\nu    $&  3 & 0.02      &   27.04 &    &           &         &   &           &          \\
\hline 
\multirow{2}{*}{$\w_1$}
&$\Omega         $&  1 & 7991.22   &   27.04 &  1 &  8008.98  &   26.85 & 1 &  7991.22  &    27.04 \\
&$\Omega +\nu    $&  2 & 0.04      &   27.04 & 15 &     0.03  &   19.72 & 2 &     0.04  &    27.04 \\
\hline 
\multirow{4}{*}{$\w_2$}
&$\Omega         $&  1 & 28903.02  & -152.96 &  1 & 28848.69  & -153.15 & 1 & 28922.58  &  -152.96 \\
&$\Omega +\nu    $&  2 & 987.12    &   27.04 &  2 &   924.23  &   18.39 & 2 &  1001.81  &    27.04 \\
&$\Omega -\nu    $&  3 & 4.87      & -152.96 & 53 &     1.48  & -146.03 &   &           &          \\
&$\Omega +2\nu   $&  4 & 0.01      &   27.04 &    &           &         &   &           &          \\
\hline
  \end{tabular}
  \llabel{Tabd}
\end{center}
\end{table*}
}

\newcommand\tabe{
\begin{table}
\begin{center}
\caption{Fundamental frequencies of the two systems. $\Omega$ and $\nu$
are the precession and nutation frequencies respectively. $\omega$ and
$n$ correspond to the precession of the periastre and the mean motion.
$\hat{g}_1$ and $\hat{l}_1$ on the one hand, and $\hat{g}_2$ and 
$\hat{l}_2$ on the other hand, are associated to
the Andoyer angles. $\hat{\psi}_2$ and $\hat{\theta}_2$ are the horizontal and
vertical libration frequencies in the resonant system $I\!I$.}
\begin{tabular}{lrr}
\hline
frequency        & \multicolumn{2}{c}{value (rad/day)} \\
                 & system $I$ & system $I\!I$ \\
\hline
$\Omega$         & -0.0312 & -0.0713 \\
$\nu$            & -0.9788 & -4.7488 \\
$\hat{\omega}$   &  0.0681 & -0.0902 \\
$n$              &  8.0052 &  9.0503 \\
$\hat{g}_1$      & 58.9763 & 63.3416 \\
$\hat{l}_1$      & -4.4062 & -8.7218 \\
$\hat{g}_2$      & 39.9703 & --      \\
$\hat{l}_2$      &-13.9042 & --      \\
$\hat{\psi}_2$   & --      &  7.5914 \\
$\hat{\theta}_2$ & --      &  4.1475 \\
\hline
\end{tabular}
\llabel{Tabe}
\end{center}
\end{table}
}

\newcommand\tabf{
\begin{table}
\begin{center}
\caption{Secular frequencies. Comparison between the integration of the
full Hamiltonian, the integration of the averaged Hamiltonian and the
analytical approximations.}
\begin{tabular}{clrr}
\hline
 &              &  $\Omega$ &     $\nu$  \\
system & type   & (rad/day) & (rad/day)  \\ \hline
\multirow{3}{*}{syst. $I$} 
 &  full        &  -0.0312  &   -0.9788  \\
 &  averaged    &  -0.0310  &   -1.1276  \\
 &  calculation &  -0.0310  &   -1.1091  \\ \hline
\multirow{3}{*}{syst. $I\!I$} 
 &  full        &  -0.0713 &   -4.7488 \\
 &  averaged    &  -0.0710 &   -3.5982 \\
 &  calculation &  -0.0710 &   -3.5629 \\ \hline
\end{tabular}
\llabel{Tabf}
\end{center}
\end{table}
}

\title{Spin axis evolution of two interacting bodies. }
\author{G. Bou\'e and J. Laskar\\[0.5cm]
\it
\small
Astronomie et Syst\`emes Dynamiques, IMCCE-CNRS UMR8028,\\
Observatoire de Paris,
77 Av. Denfert-Rochereau, 75014 Paris, France\\
 }

\date{\today}
\begin{document}
\maketitle
\icar{\vspace{10cm}}
\icar{PAGES: 66\ ; \hspace{2cm} TABLES: 8 \ ; \hspace{2cm} FIGURES: 8}
\icar{\clearpage}
\icar{
{\large Running Title : Spin axis evolution of two interacting bodies.} \\[2cm]
Corresponding author : Gwena\"el Bou\'e \\
Astronomie et Syst\`emes Dynamiques, IMCCE-CNRS UMR8028, \\
Observatoire de Paris,
77 Av. Denfert-Rochereau, 75014 Paris, France\\
email : boue@imcce.fr
\clearpage
}
\begin{abstract}
We consider the solid-solid interactions in the two body problem. The
relative equilibria have been previously studied analytically and 
general motions were numerically analyzed using some expansion of the
gravitational potential up to the second order, but only when there
are no direct interactions between the orientation of the bodies. Here we
expand the potential up to the fourth order and we show that the secular
problem obtained after averaging over fast angles, as for the precession
model of Bou\'e and Laskar [Bou\'e, G., Laskar, J., 2006. Icarus 185, 312--330]
, is integrable, but not trivially. We
describe the general features of the motions and we provide explicit
analytical approximations for the solutions. We demonstrate that the general solution 
of the secular system can be decomposed as a uniform precession 
around the total angular momentum
and a periodic symmetric orbit in the precessing frame.
More generally, we show that 
for a general  $n$-body 
system of rigid bodies in gravitational interaction, 
the regular  quasiperiodic solutions can be decomposed into a uniform 
precession around the total angular momentum, 
and a  quasiperiodic motion with one frequency less in the precessing frame.

\end{abstract}
\prep{\footnotetext[1]{E-mail address: boue@imcce.fr}}
\icar{\clearpage}

\section{Introduction}
\llabel{sec.1}
We consider here two rigid bodies orbiting each other. The main purpose
of this work is to determine the long term evolution of their spin
orientation and to a lower extent, the orientation of the orbital plane.
Examples of such systems are   binary asteroids or  a planet with a
massive satellite.

If the two bodies are spherical, then the translational and the
rotational motions are independent (e.g. Duboshin, 1958). In that case, the
orbit is purely keplerian and the proper rotation of the bodies are
uniform. General problems with triaxial bodies are more complicated,
and usually non integrable.
Even formal expansions of the gravitational potential or the proof of
their convergence can be an issue (Borderies, 1978; Paul, 1988; Tricarico,
2008). In some cases, especially for slow rotations close to 
low order spin-orbit resonances, 
the spin evolution of
rigid bodies of irregular shape  can be strongly chaotic
(Wisdom \etal, 1984; Wisdom, 1987), but we will not consider this 
situation in the present paper where we focus on 
regular and quasiperiodic motions.

Stationary solutions of spin evolution are known in the case of 
a triaxial satellite orbiting a central spherical planet (Abul'naga and 
Barkin, 1979). In their paper, Abul'naga and Barkin used canonical
coordinates, based on the Euler angles, to set the orientation of the 
satellite. On the contrary, in 1991, Wang \etal\ also studied relative
equilibria but with a vectorial approach that enabled them to 
analyze easily the stability of those solutions. For a review of
different formalisms that can be used in rigid body problems, see
(Borisov and Mamaev, 2005).

The vectorial approach turned out to be also powerful for the study of
relative equilibria of two triaxial bodies orbiting each other 
(Maciejewski, 1995). General motions of this problem were studied 
by Fahnestock and Scheeres in 2008
(hereafter FS08) in the case of the typical
binary asteroid system called 1999 KW4. For that, the authors
expanded the gravitational potential up to the second order only. In this
approximation, there is no direct interaction between the orientation of
the two bodies. Ashenberg gave in 2007 the expression of the gravitational
potential expanded up to the fourth order but didn't study the
solutions.

In (Bou\'e and Laskar, 2006) (hereafter BL06) we gave a new method to study 
the long term evolution of solid body orientations in the case of a
star-planet-satellite problem where only the planet is assumed to be
rigid. This method used a similar vectorial
approach as Wang \etal\ (1991), plus some averaging over the fast angles. 
We showed that the secular evolution of this system is 
integrable and provided the general solution. 

In the present paper, we show that the problem of two triaxial bodies orbiting
each other is very similar to the star-planet-satellite problem and thus
can be treated in the same way.

In the section 2, we compute the Hamiltonian governing the evolution of
two interacting rigid bodies. The gravitational potential is
expanded up to the fourth order and averaged over fast angles. The
resulting secular Hamiltonian is a function of three vectors only: the
orbital angular momentum and the angular momenta of the two bodies.

In a next step (section 3), we show that the secular problem is
integrable but not trivially (i.e.
it cannot be reduced to a scalar first order differential equation that can be
integrated by quadrature).
The general solution is the product
of a uniform rotation of the three vectors (global precession around the
total angular momentum) by a periodic motion (nutation). We prove also
that in a frame rotating with the precession frequency, the nutation
loops described by the three vectors are all symmetric with respect to a
same plane containing the total angular momentum.
We then derive analytical approximations of the two frequencies of the
secular problem with their amplitudes. These formulas need averaged
quantities that can be computed recursively. However we found that the
first iteration already gives satisfactory results.

In section \ref{sec.back}, we consider the general case of a $n$-body 
system of rigid bodies in gravitational interaction, 
and we demonstrate that the regular quasiperiodic solutions 
of these systems can, in a similar way,  be decomposed into a uniform 
precession, and a quasiperiodic motion in the precessing frame.

Finally, we compare our results with those of FS08 on the typical binary
asteroid system 1999 KW4. We show that their analytical expression of
the precession frequency corresponds to the simple case of a point mass
orbiting an oblate body treated in BL06. We then integrate numerically
from the full Hamiltonian, an example of a doubly asynchronous system where the
FS08 expression of the precession frequency does not apply. We compare
the results with the output of the averaged Hamiltonian and with our
numerical approximation and show that they are in good agreement.

\section{Fundamental equations}
\llabel{sec.2}

We are considering a two rigid body problem in which the interaction is
purely gravitational with no dissipative effects. Let $m_1$ and $m_2$
be the masses of the two solids. Hereafter the mass $m_2$ is called the
satellite or the secondary and the mass $m_1$ the primary. It should be
stressed that this notation does not imply any constraint on the ratio
of the masses which can even be equal to one.

The configuration of the system is described by the position vector $\rbl$ 
of the satellite barycenter relative to the primary barycenter and their
orientation expressed in an invariant reference frame. The orientations
are given by the coordinates of the principal axes $(\bI_1, \bJ_1, \bK_1)$
and $(\bI_2, \bJ_2, \bK_2)$ in which the two inertia tensors,
respectively $\In_1$ and $\In_2$, of the primary and of the secondary are
diagonal $[\In_1={\rm diag}(A_1, B_1, C_1)$ and 
$\In_2={\rm diag}(A_2, B_2, C_2)]$.

The Hamiltonian of this problem can be split into
\be
\Ham = \Ht + \He + \Hi
\llabel{eq.Hfull}
\ee
where $\Ht$ is the Hamiltonian of the free translation of the reduced
point mass $\beta=m_1m_2/(m_1+m_2)$, $\He$ describes the free rigid
rotation of the two bodies and $\Hi$ contains the gravitational
interaction.

The Hamiltonian of the free point mass is
\be
\Ht = \frac{\tr^2}{2\beta}
\llabel{eq.HT}
\ee
where $\tr=\beta\dot{\rbl}$ is the conjugate momentum of $\rbl$.

Let $\bG_1$ and $\bG_2$ be respectively the angular momentum of the
primary and of the satellite. The Hamiltonian of the free rotation is
\be
\He = \frac{\trans{\bG_1}\In_1^{-1}\bG_1}{2}
     +\frac{\trans{\bG_2}\In_2^{-1}\bG_2}{2},
\ee
where the superscript ${}^t$ in $\trans{\bf x}$ or $\trans{A}$ denotes the 
transpose of any vector $\bf x$ or matrix $A$.
It can be expressed in terms of the principal bases of the two bodies as
follows
\begin{multline}
\He = \frac{(\dop{\bG_1}{\bI_1})^2}{2A_1}
    + \frac{(\dop{\bG_1}{\bJ_1})^2}{2B_1}
    + \frac{(\dop{\bG_1}{\bK_1})^2}{2C_1} \\
    + \frac{(\dop{\bG_2}{\bI_2})^2}{2A_2}
    + \frac{(\dop{\bG_2}{\bJ_2})^2}{2B_2}
    + \frac{(\dop{\bG_2}{\bK_2})^2}{2C_2}.
\end{multline}

The interaction between the two solid bodies is the following double
integral
\be
\Hi = -\iint \frac{\cG\,dm_1\,dm_2}{\norm{\rbl+\rbl_2-\rbl_1}}
\llabel{eq.Hi}
\ee
where $\rbl_1$ and $\rbl_2$ are respectively computed relative to the primary and
satellite barycenters (cf Fig.~\ref{Figa}) and describe the two volumes. 
This part of the Hamiltonian can be
expanded in terms of Legendre polynomials and will be written as a
function of $(\rbl, \bI_1, \bJ_1, \bK_1, \bI_2, \bJ_2, \bK_2)$ in the
section~\ref{sec.potential}. 

\prep{\figa}

\subsection{Equations of motion}
\llabel{sec.motion}
The full Hamiltonian is written in the non-canonical coordinates
$(\rbl,\tr, \bI_1, \bJ_1, \bK_1, \bG_1, \bI_2, \bJ_2, \bK_2, \bG_2)$. Thus,
although the
components $(\rbl, \tr)$ keep the standard symplectic structure, $(\bI_1,
\bJ_1, \bK_1, \bG_1)$ on the one hand and $(\bI_2, \bJ_2, \bK_2, \bG_2)$
on the other hand possess the Euler-Poisson structure which leads to the
following equations of motion (Borisov and Mamaev, 2005)
\be
\begin{array}{l}
\dot{\rbl} = \grad{\tr}\Ham, \quad
{\bf\dot{\tr}} = -\grad{\rbl}\Ham \\ \\
\dot{\bG} = \grad{\bI}\Ham\times\bI +
 \grad{\bJ}\Ham\times\bJ +
 \grad{\bK}\Ham\times\bK + \grad{\bG}\Ham\times\bG  \\ \\
\dot{\bI} = \grad{\bG}\Ham\times\bI, \quad
\dot{\bJ} = \grad{\bG}\Ham\times\bJ, \quad
\dot{\bK} = \grad{\bG}\Ham\times\bK.
\llabel{eq.rot}
\end{array}
\ee
We choose these non-canonical coordinates instead of symplectic ones
because of the simplicity of the resulting equations which already 
resemble equations of precession.

\subsection{First simplification}
In the previous paragraphs, the Hamiltonian contains 
the three vectors of the principal frame $(\bI, \bJ, \bK)$ of each
body. Nevertheless, only two vectors per solid are necessary insofar as the
third can be expressed as the wedge product of the other two. We choose
to keep $\bI$ and $\bK$.

The Hamiltonian of the free rotation of the two rigid bodies can be
rewritten as follows
\begin{multline}
\He = \frac{\bG_1^2}{2B_1} + \frac{\bG_2^2}{2B_2}  \\
    + \left(\frac{1}{A_1}-\frac{1}{B_1}\right)\frac{(\dop{\bG_1}{\bI_1})^2}{2}
    + \left(\frac{1}{C_1}-\frac{1}{B_1}\right)\frac{(\dop{\bG_1}{\bK_1})^2}{2} \\
    + \left(\frac{1}{A_2}-\frac{1}{B_2}\right)\frac{(\dop{\bG_2}{\bI_2})^2}{2}
    + \left(\frac{1}{C_2}-\frac{1}{B_2}\right)\frac{(\dop{\bG_2}{\bK_2})^2}{2}.
\llabel{eq.HE}
\end{multline}

\subsection{Gravitational potential}
\llabel{sec.potential}
The distance between the two bodies is assumed to be large in comparison to their
size. Thus, in the expression of the gravitational potential (\ref{eq.Hi}), 
$\rho_1=\norm{\rbl_1}/\norm{\rbl}$ 
and $\rho_2=\norm{\rbl_2}/\norm{\rbl}$ are two small parameters.
It can then be expanded in terms of Legendre polynomials (see Appendix
A).
As it is shown below (equation~\ref{eq.HIb}), the expansion up to the
second order does not contain any interaction due to the relative
orientation of the bodies. We thus choose to expand the gravitational
potential up to the fourth order. In the computation appear integrals
such as $\int
r_i^2\,dm_i$ or $\int \rbl_i\trans{\rbl_i}\,dm_i$, $i=1,2$ 
which can be expressed in terms of moments of inertia
\be
\EQM{
\int r_i^2\,dm_i &=& &\frac{A_i+B_i+C_i}{2} \ ;\cr
\int \rbl_i\trans{\rbl_i}\,dm_i &=& &\frac{A_i-B_i+C_i}{2}\Id \cr
      &&&+ (B_i-A_i){\bI_i}\trans{\bI_i} + (B_i-C_i){\bK_i}\trans{\bK_i} \ ,
}
\ee
with $\Id$ beeing the identity matrix in ${\mathbb R}^3$.
But higher degree integrals such as $\int r_i^4\,dm_i$ also appear.
To compute these integrals, one needs more information about the bodies.
However, moments of inertia are already hardly known, at least for satellites. 
It is thus not relevant to add new unconstrained parameters. But such
integrals of inertia can be expressed as functions of $A_i$, $B_i$, $C_i$
assuming that the bodies are homogeneous ellipsoids. 
Indeed, let $(x_i,y_i,z_i)$ be the coordinates in the principal frame
of a running point of the body $i$, and
$I_{p,q,r;i}=\int x_i^py_i^qz_i^r\,dm_i$ be its integrals of inertia.
Because of the three symmetry planes of homogeneous ellipsoids, $I_{p,q,r;i}$
vanishes whenever one of $p$, $q$, $r$ is odd. Thus all the third order
integrals of inertia cancel, and the only non zero
fourth order integrals of inertia are (see Appendix B)
\be
\EQM{
\int x_i^4\,dm_i &=& \frac{15}{28m_i}(-A_i+B_i+C_i)^2 \ ; \cr
\int y_i^4\,dm_i &=& \frac{15}{28m_i}(A_i-B_i+C_i)^2 \ ; \cr
\int z_i^4\,dm_i &=& \frac{15}{28m_i}(A_i+B_i-C_i)^2 \ ; \cr
\int y_i^2z_i^2\,dm_i &=& \frac{5}{28m_i}(A_i-B_i+C_i)(A_i+B_i-C_i) \ ; \cr
\int z_i^2x_i^2\,dm_i &=& \frac{5}{28m_i}(A_i+B_i-C_i)(-A_i+B_i+C_i) \ ; \cr
\int x_i^2y_i^2\,dm_i &=& \frac{5}{28m_i}(-A_i+B_i+C_i)(A_i-B_i+C_i) \ .
}
\llabel{eq.XYZPQR}
\ee
In search of generality, we now forget the assumption of homogeneous
ellipsoids. We only keep the symmetry plane hypothesis that cancels odd
integrals. Setting
\be
\EQM{
X_i=\int x_i^4\,dm_i &\quad& P_i=\int y_i^2z_i^2\,dm_i \ ; \cr
Y_i=\int y_i^4\,dm_i &\quad& Q_i=\int z_i^2x_i^2\,dm_i \ ; \cr
Z_i=\int z_i^4\,dm_i &\quad& R_i=\int x_i^2y_i^2\,dm_i
}
\ee
the integrals appearing in the expansion of the gravitational potential
become
\be
\EQM{
\int r_i^4\,dm_i &=& 
      X_i+Y_i+Z_i+2P_i+2Q_i+2R_i \ ; \crm
\int (\dop{\s}{\rbl_i})^4\,dm_i &=&
      Y_i\s^4 + (X_i+Y_i-6R_i)(\dop{\s}{\bI_i})^4 \cr 
   && + (Z_i+Y_i-6P_i)(\dop{\s}{\bK_i})^4 \crm 
   && + 2\s^2[(3R_i-Y_i)(\dop{\s}{\bI_i})^2
      + (3P_i-Y_i)(\dop{\s}{\bK_i})^2] \crm
   && + 2[Y_i-3(P_i-Q_i+R_i)](\dop{\s}{\bI_i})^2(\dop{\s}{\bK_i})^2\ ; \crm
\int r_i^2\rbl_i\trans{\rbl_i}\,dm_i &=& 
    (Y_i+R_i+P_i)\Id + (X_i-Y_i+Q_i-P_i)\bI_i\trans{\bI_i} \cr
    && + (Z_i-Y_i+Q_i-R_i)\bK_i\trans{\bK_i}
}
\ee
where $\s$ is any vector and $i=1,2$.

With these results, the expansion of the potential gives the zeroth
order term
\be
\Hi^{(0)} = -\frac{\mu\beta}{r}
\llabel{eq.HIa}
\ee
where $\mu=\cG(m_1+m_2)$. This is the well known gravitational
interaction between two point masses. The second order terms expression
is classical and given by
\begin{multline}
\Hi^{(2)} = -\frac{1}{2}\frac{\cG}{r^3} 
    \left[ m_1(A_2-2B_2+C_2) + m_2(A_1-2B_1+C_1)\right] \\
 -\frac{3}{2}\frac{\cG m_1}{r^3} \left[ 
 (B_2-A_2)\left(\dop{\er}{\bI_2}\right)^2
+(B_2-C_2)\left(\dop{\er}{\bK_2}\right)^2 \right] \\
 -\frac{3}{2}\frac{\cG m_2}{r^3} \left[ 
 (B_1-A_1)\left(\dop{\er}{\bI_1}\right)^2
+(B_1-C_1)\left(\dop{\er}{\bK_1}\right)^2 \right]
\llabel{eq.HIb}
\end{multline}
where $\er=\rbl/r$ is the direction vector of $r$.
As mentioned before, this expression does not contain body-body
interactions but only spin-orbit ones such as $(\dop{\er}{\bK_1})^2$ or
$(\dop{\er}{\bK_2})^2$. The fourth order terms expression is
given in (\ref{eq.HIc}). In contrast to the second order terms, among the fourth order terms
there are direct interactions between the two orientations such as
$(\dop{\bK_1}{\bK_2})^2$. A similar expression was published recently in
(Ashenberg,  2007). Although  more terms 
are present in Ashenberg's paper because we have made here 
the additional assumption of symmetry of the rigid bodies, 
we could compare our expression successfully with 
the one of Ashenberg, except for a difference  in a coefficient 
that may come from a misprint in Ashenberg's paper\footnote{In (Ashenberg, 2007), 
there is a misprint in the
expression of $V_{BB'}^{(4)}$, Eq. (20). The coefficient $-3\cG/(4r^5)$
in Eq. (\ref{eq.HIc}) of the current paper corresponds to a 
coefficient $-G/(8R^5)$ in Ashenberg's notations whereas it is written
$-G/(5R^5)$ in (Ashenberg, 2007).}.

{
\icar{\small}
\begin{multline}
\Hi^{(4)} = -\frac{3}{4}\frac{\cG}{r^5} \Bigg\{ 
  (A_2-2B_2+C_2)(A_1-2B_1+C_1) \bhs \\
+\frac{1}{2}m_2 [X_1+\frac{8}{3}Y_1+Z_1-8P_1+2Q_1-8R_1] \bhs \\
+\frac{1}{2}m_1 [X_2+\frac{8}{3}Y_2+Z_2-8P_2+2Q_2-8R_2] \bhs \\
+2(B_1-A_1)(B_2-A_2)(\dop{\bI_1}{\bI_2})^2 \bhs \\
+2(B_1-A_1)(B_2-C_2)(\dop{\bI_1}{\bK_2})^2 \bhs \\
+2(B_1-C_1)(B_2-A_2)(\dop{\bK_1}{\bI_2})^2 \bhs \\
+2(B_1-C_1)(B_2-C_2)(\dop{\bK_1}{\bK_2})^2 \bhs \\
+\Big[5(A_2-2B_2+C_2)(B_1-A_1) \bhs \\
 -m_2(5X_1+\frac{20}{3}Y_1-5P_1+5Q_1-35R_1)\Big]
\left(\dop{\er}{\bI_1}\right)^2 \\
+\Big[5(A_2-2B_2+C_2)(B_1-C_1) \bhs \\
 -m_2(5Z_1+\frac{20}{3}Y_1-35P_1+5Q_1-5R_1)\Big]
\left(\dop{\er}{\bK_1}\right)^2 \\
+\Big[5(A_1-2B_1+C_1)(B_2-A_2) \bhs \\
 -m_1(5X_2+\frac{20}{3}Y_2-5P_2+5Q_2-35R_2)\Big]
\left(\dop{\er}{\bI_2}\right)^2 \\
+\Big[5(A_1-2B_1+C_1)(B_2-C_2) \bhs \\
 -m_1(5Z_2+\frac{20}{3}Y_2-35P_2+5Q_2-5R_2)\Big]
\left(\dop{\er}{\bK_2}\right)^2 \\
-20\Big[(B_1-A_1)(\dop{\er}{\bI_1})\bI_1
       +(B_1-C_1)(\dop{\er}{\bK_1})\bK_1\Big] \bhs \\
\times \Big[(B_2-A_2)(\dop{\er}{\bI_2})\bI_2
       +(B_2-C_2)(\dop{\er}{\bK_2})\bK_2\Big] \\
+\frac{35}{6}m_2\Big[
        (X_1+Y_1-6R_1)(\dop{\er}{\bI_1})^4
       +(Z_1+Y_1-6P_1)(\dop{\er}{\bK_1})^4 \\
       +2(Y_1-3P_1+3Q_1-3R_1)(\dop{\er}{\bI_1})^2(\dop{\er}{\bK_1})^2
\Big] \\
+\frac{35}{6}m_1\Big[
        (X_2+Y_2-6R_2)(\dop{\er}{\bI_2})^4
       +(Z_2+Y_2-6P_2)(\dop{\er}{\bK_2})^4 \\
       +2(Y_2-3P_2+3Q_2-3R_2)(\dop{\er}{\bI_2})^2(\dop{\er}{\bK_2})^2
\Big] \\
+35\Big[(B_1-A_1)(\dop{\er}{\bI_1})^2
       +(B_1-C_1)(\dop{\er}{\bK_1})^2\Big] \bhs \\
\times \Big[(B_2-A_2)(\dop{\er}{\bI_2})^2
       +(B_2-C_2)(\dop{\er}{\bK_2})^2\Big] \Bigg\}.
\llabel{eq.HIc}
\end{multline}
\icar{\normalsize}
}

The full Hamiltonian (\ref{eq.HE}, \ref{eq.HT}, \ref{eq.HIa}, \ref{eq.HIb} and 
\ref{eq.HIc}) together with the equations of motion (cf
section~\ref{sec.motion}) enable the integration of the system. The evolution 
of this system contains fast motions like the rotation of each body
around their axis or the orbital revolution. In comparison, the two spin
axes as well as the orientation of the orbital plane undergo secular evolutions. 
In the following, fast motions are averaged in the purpose of studying 
the long term evolution only.

\subsection{Averaging}
In this section, we average the Hamiltonian independently over all fast
angles: proper rotations and orbital motion. Although this method is
strictly valid for non resonant cases only, we will show (in
section~\ref{sec.res}) an application to a typical
primary-asynchronous, secondary-synchronous binary asteroid system where 
the motion is regular. The method still gives very acceptable results.
In the following, we forget the subscripts $1$ and $2$ whenever we
consider any of the two bodies without distinction.

\prep{\figb}

To average over proper rotations, Andoyer variables
$(G, H, L, g, h, l)$ as described in Fig.~(\ref{Figb})
are well suited. In a first step, the dependency of the full Hamiltonian
on $\bI_1$ and
$\bI_2$ is removed by averaging over $l_1$ and $l_2$. We have
\be
\bI = \bpm \cos l \\ \sin l \\ 0 \epm_{(\n, \n', \bK)}
\ee
where $\n$ is defined in Fig.~(\ref{Figb}) and $\n'=\bK\times\n$. The
vectors $\n$, $\n'$ and $\bK$ are independent of $l$, thus
\be
\begin{array}{lcl}
\ds \LA \bI \RA_l &=& \ds {\bf 0} \ ; \\ \\
\ds \LA \bI\trans{\bI} \RA_l &=& \ds \frac{1}{2}(\Id-\bK\trans{\bK}) \ ; \\ \\
\ds \LA (\dop{\s}{\bI})^4 \RA_l &=& \ds 
         \frac{3}{8}\left[\s^2-(\dop{\s}{\bK})^2\right]^2 \ ,
\end{array}
\llabel{eq.mI}
\ee
where $\s$ is again any vector. After this averaging, the Hamiltonian of
the free rotation becomes
\begin{multline}
\LA\He\RA_{l_1,l_2} = \frac{\bG_1^2}{2\fA_1}
+\left(\frac{1}{C_1}-\frac{1}{\fA_1}\right)
\frac{(\dop{\bG_1}{\bK_1})^2}{2} \\
  +   \frac{\bG_2^2}{2\fA_2}
+\left(\frac{1}{C_2}-\frac{1}{\fA_2}\right)
\frac{(\dop{\bG_2}{\bK_2})^2}{2}
\end{multline}
where 
\be
\frac{1}{\fA} = \frac{1}{2}\left(\frac{1}{A}+\frac{1}{B}\right).
\llabel{eq.fA}
\ee
And the second and the fourth order terms of the interaction 
\begin{multline}
\LA\Hi^{(2)}\RA_{l_1,l_2} = -\frac{\cG \fC_1m_2}{2r^3}
\left[1-3(\dop{\er}{\bK_1})^2\right] \\
            -\frac{\cG \fC_2m_1}{2r^3}
\left[1-3(\dop{\er}{\bK_2})^2\right],
\llabel{eq.Hi1}
\end{multline}
\be
\EQM{
\LA\Hi^{(4)}\RA_{l_1,l_2} &=& 
   -\frac{3}{8}\frac{\cG m_2\fT_1}{r^5}
    \left[1-10(\dop{\er}{\bK_1})^2+\frac{35}{3}(\dop{\er}{\bK_1})^4\right] \crm
&& -\frac{3}{8}\frac{\cG m_1\fT_2}{r^5}
    \left[1-10(\dop{\er}{\bK_2})^2+\frac{35}{3}(\dop{\er}{\bK_2})^4\right] \crm
&& -\frac{3}{4}\frac{\cG\fC_1\fC_2}{r^5} \Big[ 1
   + 2(\dop{\bK_1}{\bK_2})^2 
   - 5(\dop{\er}{\bK_1})^2 \crm
&& - 5(\dop{\er}{\bK_2})^2 
   -20(\dop{\er}{\bK_1})(\dop{\er}{\bK_2})(\dop{\bK_1}{\bK_2}) \crm
&& +35(\dop{\er}{\bK_1})^2(\dop{\er}{\bK_2})^2
\Big] \ ,
}
\llabel{eq.Hi2}
\ee
where $\fC = C-(A+B)/2$ and 
\be
\fT=\frac{3}{8}(X+Y)+Z-3(P+Q)+\frac{3}{4}R.
\llabel{eq.fTXYZPQR}
\ee

In a next step, the averaging over the angle $g$ is performed.
This corresponds to the averaging of $\bK$ around $\w=\bG/G$ (cf
Fig.~\ref{Figb}). Indeed, in the general case the angular momentum $\bG$
is not aligned with the axis of maximum inertia $\bK$, which is
implicitly assumed in the gyroscopic approximation. Instead, if there is
an angle $J$ between these two vectors then
\be
\bK = \bpm \sin J\sin g \\ -\sin J\cos g \\ \cos J \epm_{(\N_1, \U_1, \w)}
\ee
where $\N_1$ is defined in Fig.~\ref{Figb} and $\U_1=\w\times\N_1$. The
vectors $\N_1$, $\U_1$ and $\w$ are independent of $g$, so
\be
\begin{array}{lcl}
\ds \LA \bK \RA_g &=& \ds (\cos J) \w \ ; \\ \\
\ds \LA \bK\trans{\bK} \RA_g &=& \ds 
         \frac{1}{2}(\sin^2J)\Id + \left(1-\frac{3}{2}\sin^2J\right)\w\trans{\w} \ ; \\ \\
\ds \LA (\dop{\s}{\bK})^4 \RA_g &=& \ds 
        \left(1-5\sin^2J+\frac{35}{8}\sin^4J\right)(\dop{\s}{\w})^4 \\ \\
&&  \ds +3\sin^2J\left(1-\frac{5}{4}\sin^2J\right)\s^2(\dop{\s}{\w})^2 \\ \\
&&  \ds +\frac{3}{8}\sin^4J\s^4 \ ,
\end{array}
\ee
where $\s$ is any vector. After averaging over $g_1$ and $g_2$, the
conjugated momenta $G_1$ and $G_2$ become constant. The averaged 
Euler Hamiltonian which depends only on $G_1$ and $G_2$
\be
\EQM{
\LA\He\RA_{l,g} &=& \left(
\frac{\cos^2J_1}{C_1}+\frac{\sin^2J_1}{\fA_1}
\right)\frac{\bG_1^2}{2} \crm &&
+ \left(
\frac{\cos^2J_2}{C_2}+\frac{\sin^2J_2}{\fA_2}
\right)\frac{\bG_2^2}{2} 
}
\ee
is now a constant and can be ignored. In this expression, 
$\fA$ is still the harmonic mean
of $A$ and $B$ (\ref{eq.fA}). The only change in 
the interaction is the substitution
of $\fC$ and $\fT$ in (\ref{eq.Hi1}-\ref{eq.Hi2}) by
\be
\begin{array}{lcl}
\ds\fC' &\ds=& \ds\left(1-\frac{3}{2}\sin^2J\right)\fC \\ \\
\ds\fT' &\ds=& \ds\left(1-5\sin^2J+\frac{35}{8}\sin^4J\right)\fT
\llabel{eq.fT}
\end{array}
\ee
and $(\bK_1,\bK_2)$ by $(\w_1,\w_2)$. For fast rotating non-rigid bodies,
the angle $J$ is assumed to be very small as a result of internal
dissipation ($J\approx10^{-7}$ radians for the Earth). In that case, the
gyroscopic approximation $J=0$ is a good approximation since the correction
obtained after averaging over fast angles is in $O(\sin^2 J)$.
Nevertheless, for slow rotating triaxial asteroids, the angle $J$ may be large
and the gyroscopic approximation may not be valid.

In a third step the Hamiltonian is averaged over the orbital motion. First over the
mean anomaly $M$, and then over the longitude of periapse $\omega$. 
The first average is computed using the formulas of the Appendix~C
and for the second one, we have similar equations as (\ref{eq.mI})
\be
\begin{array}{lcl}
\ds \LA \bI \RA_\omega &=& \ds {\bf 0} \ ; \\ \\
\ds \LA \bI\trans{\bI} \RA_\omega &=& \ds \frac{1}{2}(\Id-\w\trans{\w}) \ ; \\ \\
\ds \LA (\dop{\s}{\bI})^4 \RA_\omega &=& \ds 
         \frac{3}{8}\left[\s^2-(\dop{\s}{\w})^2\right]^2 \ ,
\end{array}
\ee
where $\bI$ now denotes direction of the periapse and $\w$ the normal of the
orbit. The resulting secular Hamiltonian $\Hs =
\LA\Ham\RA_{l_1,l_2,g_1,g_2,M,\omega}$ is thus
\be
\EQM{
\Hs &=& 
      \left(\frac{\cos^2J_1}{C_1}+\frac{\sin^2J_1}{\fA_1}
      \right)\frac{\bG_1^2}{2}
   +  \left(\frac{\cos^2J_2}{C_2}+\frac{\sin^2J_2}{\fA_2}
      \right)\frac{\bG_2^2}{2} \cr \cr
&& - \frac{\mu\beta}{2a}  \cr \cr
&& +\frac{\cG}{4a^3(1-e^2)^{3/2}}
    \left[m_2\fC_1^\prime(1-3x^2) + 
          m_1\fC_2^\prime(1-3y^2)\right] \cr \cr
&& -\frac{9}{32}\frac{\cG}{a^5(1-e^2)^{7/2}}
   \left(1+\frac{3}{2}e^2\right)\Bigg[ \cr \cr
&& \fC_1^\prime\fC_2^\prime(1-5x^2-5y^2+2z^2-20xyz+35x^2y^2) \cr \cr
&& +\frac{m_2\fT_1^\prime}{2}\left(1-10x^2+\frac{35}{3}x^4\right) \cr \cr
&& +\frac{m_1\fT_2^\prime}{2}\left(1-10y^2+\frac{35}{3}y^4\right)
   \Bigg]
}
\ee
where $x=(\dop{\w}{\w_1})$, $y=(\dop{\w}{\w_2})$ and
$z=(\dop{\w_1}{\w_2})$. 
Let us write $\Hs$ in the more compact form
\be
\Hs = 
 - \frac{\fa}{2} x^2 
 - \frac{\fb}{2} y^2
 - \frac{\fc}{2} z^2
 + \fd xyz
 - \frac{\fe}{4} x^4
 - \frac{\ff}{4} y^4
 - \frac{\fg}{2} x^2y^2
 + \fh
\llabel{eq.Hs}
\ee
where
\be
\begin{array}{lcl}
\ds\fa &=& \ds\ka m_2\fCa
   - \frac{5}{2}\kb(\fCa\fCb+m_2\fTa) \cr\cr 
\ds\fb &=& \ds\ka m_1\fCb
   - \frac{5}{2}\kb(\fCa\fCb+m_1\fTb) \cr\cr
\ds\fc &=& \ds\kb\fCa\fCb \cr\cr
\ds\fd &=& \ds5\kb\fCa\fCb \cr\cr
\ds\fe &=& \ds\frac{35}{6}\kb m_2\fTa \cr\cr
\ds\ff &=& \ds\frac{35}{6}\kb m_1\fTb \cr\cr
\ds\fg &=& \ds\frac{35}{2}\kb\fCa\fCb \cr\cr
\ds\fh &=& \ds\frac{1}{6}\ka(m_2\fCa+m_1\fCb)
           -\frac{1}{8}\kb(2\fCa\fCb+m_2\fTa+m_1\fTb) \cr\cr
&&         +\ds\LA\He\RA_{l,g}-\frac{\mu\beta}{2a}
\end{array}
\llabel{eq.abcdefgh}
\ee
with
\be
\EQM{
\ka &=& \frac{3}{2}\frac{\cG}{a^3(1-e^2)^{3/2}} \cr\cr
\kb &=& \frac{9}{8}\frac{\cG}{a^5(1-e^2)^{7/2}}
      \left(1+\frac{3}{2}e^2\right).
}
\ee

\section{Secular equations}
\llabel{sec.seq}
The secular Hamiltonian $\Hs$ (\ref{eq.Hs}) is similar to the one obtained in
BL06 although its expression is slightly more complicated. 
The difference with BL06 is that the secular Hamiltonian is not anymore
the equation of an ellipsoid in $(x,y,z)$. A few results in BL06 were
proved for this special surface. 
We recall here the main steps of the derivation of the solutions adapted
to the new surface defined by the current secular Hamiltonian.

The Hamiltonian $\Hs$ is only a function of the angular momenta 
$(\bG, \bG_1, \bG_2)$. The equations of motion of these quantities are
\be
\EQM{
\dot\bG =  \grad{\bG} \Hs \times \bG \ , \crm
\dot\bG_1 =  \grad{\bG_1} \Hs \times \bG_1 \ ,\crm
\dot\bG_2 =  \grad{\bG_2} \Hs \times \bG_2 \ .\crm
}
\ee
We thus have $\dop{\bG}{\dot\bG}= \dop{\bG_1}{\dot\bG_1} = 
\dop{\bG_2}{\dot\bG_2} = 0$ which means that the norms $\gamma=\norm{\bG}$,
$\beta=\norm{\bG_1}$ and $\alpha=\norm{\bG_2}$ are constant. It is thus
possible to write the general equations directly in terms of the unit
vectors $(\w, \w_1, \w_2)$
\be
\EQM{
\dot\w =  \frac{1}{\gamma} \grad{\w} \Hs \times \w \ , \crm
\dot\w_1 =   \frac{1}{\beta}\grad{\w_1} \Hs \times \w_1 \ ,\crm
\dot\w_2 =   \frac{1}{\alpha}\grad{\w_2} \Hs \times \w_2 \ .\crm
}
\ee
From the expression of the secular Hamiltonian
(\ref{eq.Hs}), we get
\be
\EQM{
\dot\w = -\frac{p}{\gamma} \w_1 \times \w 
         -\frac{q}{\gamma} \w_2 \times \w \ , \crm
\dot\w_1 = -\frac{p}{\beta} \w \times \w_1 
           -\frac{s}{\beta} \w_2 \times \w_1 \ , \crm
\dot\w_2 = -\frac{q}{\alpha} \w \times \w_2 
           -\frac{s}{\alpha} \w_1 \times \w_2 \ , \crm
\llabel{eq.sec}
}
\ee
where
\be
\EQM{
p = \fa x-\fd yz+\fe x^3+\fg xy^2 \ ,\crm
q = \fb y-\fd xz+\ff y^3+\fg x^2y \ ,\crm
s = \fc z-\fd xy \ .\crm
}
\llabel{eq.pqs}
\ee

The problem has 9 degrees of freedom, the coordinates of $\G$, $\G_1$ and $\G_2$,
and the equations (\ref{eq.sec}-\ref{eq.pqs}) are non-linear. At
first glance the resolution is difficult. There are 7 first integrals
\be
\EQM{
\norm{\w}  = 1 \crm
\norm{\w_1}  = 1 \crm
\norm{\w_2}  = 1 \crm
\fa x^2 + \fb y^2 + \fc z^2 - 2\fd xyz + \frac{\fe}{2} x^4
+ \frac{\ff}{2} y^4 + \fg x^2y^2 = -2 \Hs \crm
\gamma \w + \beta \w_1 + \alpha \w_2 = \bW_0
}
\llabel{integrals}
\ee
where $\bW_0$ is the total angular momentum. Thus one misses one constant 
of motion to integrate the problem by quadrature. 
The next section shows how to solve the relative motion of
the three vectors that contains enough constants of motion.

\subsection{Relative solution}
In the previous section, we have shown that the number of first integrals is not
large enough to solve the full problem. But the number of degrees of
freedom can be decreased by considering only the relative distance between
the vectors. These distances are given by the dot products
$x=\dop{\w}{\w_1}$, $y=\dop{\w}{\w_2}$ and $z=\dop{\w_1}{\w_2}$.
From the equations (\ref{eq.sec}), one can derive the new equations of
motion
\be
\EQM{
\dot x = \left( -\frac{q}{\gamma} + \frac{s}{\beta}  \right) v \ ,\crm
\dot y = \left( -\frac{s}{\alpha} + \frac{p}{\gamma} \right) v \ ,\crm
\dot z = \left( -\frac{p}{\beta}  + \frac{q}{\alpha} \right) v \ .\crm
}
\llabel{eq.xyz}
\ee
where $v=\dop{(\w\times\w_1)}{\w_2}$ is the volume defined by the 3
vectors. It can be expressed in terms of $x$, $y$ and $z$ through the
Gram determinant
\be
v^2 = \left|\begin{matrix}
1 & x & y \\
x & 1 & z \\
y & z & 1 
\end{matrix}\right|
=1-x^2-y^2-z^2+2xyz.
\ee

This restricted problem has only 3 degrees of freedom and 2 first integrals
\be
\EQM{
\fa x^2 + \fb y^2 + \fc z^2 - 2\fd xyz + \frac{\fe}{2} x^4
+ \frac{\ff}{2} y^4 + \fg x^2y^2 = -2 \Hs \crm
\gamma\beta x + \alpha\gamma y + \alpha\beta z = K \crm
}
\ee
the second integral being simply derived from $2K = \norm{\bW_0}^2 -
(\gamma^2 + \beta^2 + \alpha^2)$. The motion in $(x, y, z)$ is thus
integrable and the solution evolves in the intersection $\cC$ of the quartic 
$\Hs=\const$ and the plane $K=\const$\ \footnote{In the whole paper,
$\const$ means any constant value.}.
Moreover, the evolution is limited to the interior of the 
$v^2(x,y,z)=0$ surface that will be henceforth called the Cassini
berlingot\footnote{A berlingot is a famous tetrahedron hard candy with
rounded edges.} as in BL06 (cf Fig.~\ref{Figc}). Outside
this surface we would have $v^2<0$ which is not possible (see BL06).

\prep{\figc}

\subsubsection{Shape of the quartic surface}
The constraint $\Hs=\const$ defines a quartic surface $\cQ$ in $(x, y, z)$. Quartic 
surfaces can have very different shapes, nevertheless setting 
$z'=z-\frac{\fd}{\fc}xy$, one obtains 
\be
-2\Hs = \fa x^2 + \fb y^2 + \fc z'^2 + \frac{\fe}{2} x^4
+ \frac{\ff}{2} y^4 + \left(\fg - \frac{\fd^2}{\fc} \right) x^2y^2
\llabel{eq.nHs}
\ee
which is a biquadratic. The new surface $\cQ'$ defined by
(\ref{eq.nHs}) is thus symmetric in $x$, $y$ and $z'$. In $(x^2,
y^2, z')$ the surface $\cQ'$ can be either an ellipsoid, a paraboloid
or an hyperboloid depending on the sign of 
\be
\delta = \fe\ff-\left(\fg-\frac{\fd^2}{\fc}\right)^2.
\ee
If $\delta>0$ then it is an ellipsoid and $x$, $y$, $z'$ and thus $z$ are
bounded. In the other case,  
$\cQ'$ is either an elliptic paraboloid if 
$\delta=0$ or an hyperboloid of one or two
sheets depending on the value of $\Hs$ if $\delta<0$.
Thus, $x$, $y$, $z$ are unbounded.

From the definition of the coefficients $\fa$--$\fg$ (\ref{eq.abcdefgh}), 
$\delta$ can be rewritten in the following form
\be
\delta = \left(\frac{35}{6}\right)^2\kb^2 m_1m_2\fTa\fTb - 
         \left(\frac{15}{2}\right)^2\kb^2 \fCa^2\fCb^2.
\ee
Using the definition of the coefficients $\fC'$ and $\fT'$ (\ref{eq.fT})
and (\ref{eq.fTXYZPQR}), we get
\be
\EQM{
\delta &=& \left(\frac{15}{2}\right)^2\kb^2\fC_1^2\fC_2^2\Bigg[ \crm
&&
\eta\left(1-5\sin^2J_1+\frac{35}{8}\sin^4J_1\right) \crm
&& \times    \left(1-5\sin^2J_2+\frac{35}{8}\sin^4J_2\right) \crm
&& -\left(1-\frac{3}{2}\sin^2J_1\right)^2
    \left(1-\frac{3}{2}\sin^2J_2\right)^2 
\Bigg]
}
\ee
where $\eta$ is a positive parameter related to the shapes of the rigid 
bodies
\be
\eta = \left(\frac{7}{9}\right)^2\frac{m_1\fT_1 m_2\fT_2}{\fC_1^2\fC_2^2}\ .
\ee

Let us look to the range of the possible values of $\eta$ in the case of an
homogeneous ellipsoids. We have the relation between $\fT$ and
$\fC$ given by the equations (\ref{eq.XYZPQR})
\be
\fT=\frac{15}{7m}\left[\fC^2+\frac{1}{8}(B-A)^2\right]\ .
\ee
The lowest value of $\eta$ is thus obtained for $A=B$, i.e. for
axisymmetric bodies. In that case, $\eta_{\text{min}}=25/9$. Conversely,
the largest
value of $\eta$ is attained when $(B-A)^2$ is maximal, thus when $B=C$ and
$A=0$, that is, in the limiting case where the bodies are extremly thin rods.
In this second case, $\eta_{\text{max}}=25/4$. So, for homogeneous
ellipsoids, $\eta$ is constrained between $\eta_{\text{min}}$ and
$\eta_{\text{max}}$. 

Figure \ref{Figca} shows the domains where the surface $\cal Q'$ is an
ellipsoid ($\cal E$) or an hyperboloid ($\cal H$) as a function of the
angles $J_1$ and $J_2$. The two sets of curve correspond to
$\eta=\eta_{\text{min}}$ and $\eta=\eta_{\text{max}}$. As $\delta$ is a
function of $\sin^2J_1$ and $\sin^2J_2$, the figure can be extended up
to $180$ degrees applying axial symmetry around the axis $J_1=90$
degrees and $J_2=90$ degrees.

\prep{\figca}

\subsubsection{Description of the solutions}
\llabel{sec.periodic}
In BL06, we show that when the surface $\cQ$ is an ellipsoid then
the evolution of $(x,y,z)$ presents two kinds of solutions. We have 
called {\sl special solutions} the solutions where $\cC$ is
totally included in the Cassini berlingot $\cB$. This means that the
vectors $\w$, $\w_1$ and $\w_2$ are never collinear. This happens only
when the three vectors are almost orthogonal. The second class of
solutions are the {\sl general solutions}, more frequent in
astronomical problems, for which $\cC$ crosses the Cassini berlingot (Fig.~\ref{Figd}). 
In that case $\bM=(x, y, z)$ does periodic returns inside the Cassini berlingot up to its 
surface and the volume $v$ defined by the three vectors $\w$, $\w_1$ and $\w_2$ is conserved
over one period. In both cases, solutions are periodic.

\prep{\figd}

There are also special cases that happen when the orbit of $(x,y,z)$ is 
tangent to the Cassini berlingot. At the tangency there is indeed a fixed point.
In that state, the three vectors remain in a plane that precesses in
time. It is called a Cassini state (Colombo, 1966; Peale, 1969; Ward,
1975; BL06). If an initial condition is chosen along such special orbits but
strictly outside the fixed point, then the system cannot reach the
stationary point in finite time and it is the only case where $x$, $y$, $z$ 
are not periodic.

Here, we have the same results except when the quartic $\cQ$ is 
unbounded. In that case, we cannot have special solutions.

\subsection{Global solution}
\label{sec.global}
Knowing that $x$, $y$, $z$ are periodic functions of time, it is possible
to get general features of the global motion. For that, let us rewrite the
secular equations (\ref{eq.sec}) in a new form so as to obtain a
linear differential system with periodic coefficients.

Let us assume as in BL06 that
the vectors $(\w, \w_1, \w_2)$ are not coplanar $(v\neq0)$. Let $\cW$ be
the matrix $(\w, \w_1, \w_2)$ and $V$ the Gram matrix of the basis $(\w,
\w_1, \w_2)$
\be
V = \bpm 1 & x & y \\ x & 1 & z \\ y & z & 1 \epm \ .
\ee
Using the expression of the wedge product in the basis $(\w, \w_1, \w_2)$ 
(see the appendix B of BL06),
the equations of motion (\ref{eq.sec}) can be written in the following
form
\be
\dot\cW = vV^{-1}\cW\cA.
\llabel{eq:globper}
\ee
Here we correct a mistake\footnote{In BL06, we have 
incorrectly stated that the averaged differential system (51) could 
be written as $\dot{\cal W} = {\cal W}{\cal B}$ where $\cB=vV^{-1}\cA$
is a matrix depending only on $(x,y,z)$. In fact the correct expression 
is $\dot{\cal W}=vV^{-1}{\cal W}\cA$. In BL06, the proof following the
equation (51) has to be modified. This is done in the present paper. 
The results remain identical.}
in the demonstration of the proposition
1, given in section 4 in BL06 (see the erratum Bou\'e and Laskar, 2008). 

In (\ref{eq:globper}), $vV^{-1}$ and $\cA$ are matrices depending only on $(x, y, z)$
that are periodic functions of period $T$. Indeed
\be
\cA =  \bpm  
       0          &  \frac{s}{\beta} &  -\frac{s}{\alpha} \\
-\frac{q}{\gamma} &          0       &   \frac{q}{\alpha} \\
 \frac{p}{\gamma} & -\frac{p}{\beta} &         0 
\epm
\ee
and
\be
V^{-1}= \frac{1}{v^2} \bpm 1-z^2 & yz-x & xz-y \\ yz-x & 1-y^2 & xy-z \\
xz-y & xy-z & 1-x^2 \epm \ .
\ee

Thus, if $\cW(t)$ is a solution of (\ref{eq:globper}), then $\cW(t+T)$
is also a solution. Let us denote
\be
\cR_T(t) = \cW(t+T)\cW(t)^{-1}\ .
\llabel{eq:defR}
\ee
We need to prove that $\cR_T(t)$ is constant with $t$. As the Gram
matrix $V$ of the vectors $(\w(t), \w_1(t), \w_2(t))$ is $T$-periodic,
the norm is conserved by linear transformation $\cR_T(t)$ that send
$\cW(t)$ into $\cW(t+T)$, and $\cR(t)$ is thus an isometry of $\BR^3$.
Moreover, this isometry is positive, as the volume $v$ is conserved over
a full period $T$ (see section \ref{sec.periodic}). The invariance of the total
angular momentum $\bW_0$
(\ref{integrals}) then implies that $\cR_T(t)$ is a rotation matrix of
axis $\bW_0$.

As $\cR_T(t)$ is a rotation in $\BR^3$, we have for all $\w_i, \w_j$
in $\{\w,\w_1,\w_2\}$, 
\be
\EQM{
\w_i(t+T)\times\w_j(t+T) &=& (\cR_T(t)\w_i(t))
                       \times(\cR_T(t)\w_j(t)) \cr
                         &=& \cR_T(t)(\w_i(t)\times\w_j(t))\ .
}
\ee
From the equations of motion (\ref{eq.sec}), we can thus derive
\be
\dot{\cW}(t+T) = \cR_T(t)\dot{\cW}(t)\ .
\ee
On the other hand, as $\cW(t+T) = \cR_T(t)\cW(t)$ (\ref{eq:defR}), we
deduce that for all $t$, 
\be
\dot{\cR}_T(t)\cW(t) = 0\ .
\ee
$\cR_T(t)$ is thus a constant matrix $\cR_T$. Now, let us denote $\cR(t)$ 
the rotation of axis $\bW_0$ and angle $t\theta_T/T$ (i.e. $\cR(T)=\cR_T$).
We have

\begin{prop}
The complete solution $\cW(t)$ can be expressed on
the form
\be
\cW(t) = \cR(t)\tilde{\cW}(t)\ ,
\llabel{eq.global}
\ee
where $\tilde{\cW}(t)$ is periodic with period $T$, and $\cR(t)$ a
uniform rotation of axis $\bW_0$ and angle $t\theta_T/T$. The motion has
two periods: the (usually) short period $T$ and the precession period
\be
T' = \frac{2\pi}{\theta_T}T\ .
\ee
\end{prop}

\subsection{Properties of the solution}
A more precise result on the periodic loops can be proved. But
before, one needs to write the instantaneous precession speed as a 
function of $(x,y,z)$.

\subsubsection{Instantaneous precession rate}
Let us write the time derivative of the precession angle of $\w$ as a 
function of $(x,y,z)$. 
The expressions for the other vectors can be obtained in the same
way. The following approach is highly inspired by BL06. We set
$W_0=\norm{\bW_0}$ the norm of the total angular momentum and 
$\w_0=\bW_0/W_0$ its direction vector. With
\be
\zeta = \dop{\w}{\w_0},
\llabel{eq.zeta}
\ee
the projection $\bL$ of $\w$ on the plane orthogonal to $\w_0$ is
\be
\bL = \w-\zeta\w_0.
\llabel{eq.defL}
\ee
Assuming $\w\neq\w_0$, we get $\zeta<1$. With $L=\norm{\bL}$, the expression of $\bL$ gives
\be
L=\sqrt{1-\zeta^2} \qquad{\text{and}}\qquad \dot{L} =
-\frac{\zeta\dot{\zeta}}{\sqrt{1-\zeta^2}}.
\ee
Moreover, setting $\bl=\bL/L$, we get
\be
\bL=L\bl \qquad{\text{and}}\qquad \dot{\bL}=\dot{L}\bl+\dot{\theta}(\w_0\times\bL)
\llabel{eq.dotL}
\ee
which yields to
\be
\dot{\bL}^2 = \dot{L}^2+\dot{\theta}^2(\w_0\times\bL)^2 = 
\dot{L}^2+\dot{\theta}^2(1-\zeta^2).
\ee
Now, from the expression of $\bL$ (\ref{eq.defL}), we can also write
\be
\dot{\bL}=\dot{\w}-\dot{\zeta}\w_0 \qquad{\text{and}}\qquad
\dot{\bL}^2=\dot{\w}^2-\dot{\zeta}^2.
\ee
Finally, we have
\be
\dot{\theta}^2 = \frac{\dot{\w}^2-\dot{\zeta}^2/(1-\zeta^2)}{1-\zeta^2}.
\llabel{eq.theta}
\ee
This final expression is an explicit function of $(x,y,z)$. Indeed, from
(\ref{integrals}), one has
\be
\zeta = \frac{1}{W_0}(\gamma+\beta x+\alpha y)
\llabel{eq.zetaxy}
\ee
and thus
\be
\dot{\zeta} = \frac{v}{W_0}\left(\frac{\alpha p}{\gamma}-\frac{\beta q}{\gamma}\right).
\llabel{eq.dotzeta}
\ee
We have also from (\ref{eq.sec})
\be
\dot{\w}^2 = \frac{1}{\gamma^2}\left( p^2+q^2+2pqz-(px+qy)^2 \right).
\ee
so (\ref{eq.theta}) can be written on the form
\be
\dot{\theta}^2 = \Theta(x,y,z).
\llabel{eq.Theta}
\ee
The sign of $\dot{\theta}$ can be determined
through (\ref{eq.dotL}). Indeed $\dot{\theta}$ is a function of
$(\w,\w_1\w_2)$, but its sign can only change when $\dot{\theta}=0$,
that is from (\ref{eq.theta}), when
\be
\dot{\w}^2(1-\zeta^2)=\dot{\zeta}^2.
\ee
The equation (\ref{eq.Theta}) thus gives the instantaneous precession
rate of $\w$ as a function of $x$, $y$, $z$. Same results can easily 
be obtained for the other two vectors $\w_1$ and $\w_2$.

\subsubsection{Symmetry of the nutation}
It is now possible to prove a more precise result on the periodic loops
generated by $\w$, $\w_1$ and $\w_2$ in the precessing frame. This is
the same result as in BL06 that was given for a three body problem with
only one rigid body.

\medskip

\begin{prop}
In the frame rotating uniformly with the precession
period, the three vectors $\w$, $\w_1$, $\w_2$ describe periodic
loops $\cL$, $\cL_1$, $\cL_2$ that are all symmetric with respect to the
same plane $\cS$ containing $\w_0$.
\end{prop}

\medskip

\begin{consequence}
Let us call $\cP$, $\cP_1$, $\cP_2$ the averages of
$\w$, $\w_1$, $\w_2$ over the nutation angle. $\cP$, $\cP_1$, $\cP_2$
are respectively the pole of the orbit, the pole of the spin of the
primary and the pole of the spin of the secondary. Due to the symmetry
of the loops, the three poles $\cP$, $\cP_1$ and $\cP_2$ remain in the
symmetry plane $\cS$ containing $\w_0$, and precessing uniformly around
$\w_0$. Each vector $\w$, $\w_1$, $\w_2$ nutates around its pole,
respectively $\cP$, $\cP_1$, $\cP_2$.
\end{consequence}

\medskip
 
\begin{proof}
As in BL06, we will consider uniquely $\w$, the other cases
being similar. We consider here a general solution, for which the orbit
of $(x,y,z)$ crosses the Cassini berlingot (Fig.~\ref{Figd}). We choose
the origin of time in $\tau_+$ which corresponds to an orbital angular
momentum $\w_+$.
Let $\sigma$ be the arc length described by $\bM=(x,y,z)$ computed from 
$\bM_+=\bM(\tau_+)$. From (\ref{eq.xyz}) we have
\be
\dot{\sigma} = v\sqrt{f(x,y,z)}
\ee
where
\be
f(x,y,z) = \left(\frac{q}{\gamma}-\frac{s}{\beta}\right)^2
          +\left(\frac{s}{\alpha}-\frac{p}{\gamma}\right)^2
          +\left(\frac{p}{\beta}-\frac{q}{\alpha}\right)^2.
\ee
$f(x,y,z)=0$ if and only if $\alpha p = \beta q = \gamma s$. This
condition corresponds to a fixed point of the system. Else $f(x,y,z)$
is strictly positive.

Thus $\dot{\sigma}$ is a function of $(x,y,z)$ and
has the sign of $v$. For $t<0$, the orbit in the $(x,y,z)$ describes the arc
$(\tau_-,\tau_+)$, thus $\sigma$ decreases from $\sigma_-$ down to
$\sigma_+=0$, and $v<0$. Conversely, for $t>0$ the orbit describes the
same arc in the reverse way $(\tau_+,\tau_-)$, hence $v>0$. As $x$, $y$,
$z$ are functions of the arc length $\sigma$, we can write
\be
\dot{\sigma} = \left\{
\EQM{
-F(\alpha), \quad\text{if } t<0, \cr
+F(\alpha), \quad\text{if } t>0
}\right.
\ee
where $F(\alpha) = |v|\sqrt{f(x,y,z)}$. We conclude that $\sigma$ and
thus $\bM=(x,y,z)$ are even, that is $\bM(-t)=\bM(t)$.

The rest of the proof is identical to the one of BL06. We recall it for
completeness. From (\ref{eq.Theta})
\be
\dot{\theta}^2(t) = \Theta(x,y,z),
\ee
we deduce that $\dot{\theta}^2(t)$ is even. Moreover, as the
differential system (\ref{eq.sec}) is polynomial, the solutions $\w$,
$\w_1$, $\w_2$ are analytical in time
$t$, and so will be the coordinate angle $\theta(t)$ of $\w$. The lemma
of BL06 thus implies that $\dot{\theta}(t)$ is odd or even.
If $\dot{\theta}(t)$ is even on
$[-T/2,T/2]$, for all $h\in[0,T/2]$, we have
$\theta(h)-\theta(0)=\theta(0)-\theta(-h)$. As the cosine $\zeta$ of the
angle from $\w$ and $\w_0$ (\ref{eq.zeta}) depends only on $x$, $y$
(\ref{eq.zetaxy}), we have $\zeta(h)=\zeta(-h)$, and $\w(h)$ and $\w(-h)$
are symmetrical with respect to the $(\w_0,\w_+)$ plane. It will still
be the same in the rotating frame with the precession period. In this
rotating frame, the periodic loop generated by $\w$ is thus symmetric
with respect to the plane $(\w_0,\w_+)$.

Moreover, at $t=0$ ($\tau_+$), the volume $v$ is null, and thus $\w_0$,
$\w$, $\w_1$, $\w_2$ are coplanar. In the rotating frame, all three
orbits generated by $\w$, $\w_1$, $\w_2$ are thus symmetrical with
respect to the same plane $(\w_0,\w_+)$.

The only case where $\dot{\theta}(t)$ is odd, occurs when $\dot{\theta}(0)=0$.
As $v(0)=0$, we have $\dot{\zeta}(0)=\bf 0$ (\ref{eq.dotzeta}) and $\dot{\w}=\bf 0$ 
(\ref{eq.theta}). In the same way, we have $\dot{\w}_1(0)=\dot{\w}_2(0)=\bf 0$, 
and the vector field (\ref{eq.sec}) vanishes at $t=0$. The three vectors
$\w$, $\w_1$, $\w_2$ are thus stationary and coplanar. 
\end{proof}

This is a special Cassini state where the precession frequency is zero.

\subsection{Computation of the two periods}
The nutation period and the precession period are two key parameters of
the problem since the global solution is the product of these two
motions (\ref{eq.global}). Let us see how the values can be derived.

The three dot products $(x(t), y(t), z(t))$ are $T$-periodic where $T$
is the nutation period. This period can thus be calculated from the
expression of $(x(t), y(t), z(t))$. Given the two first integrals
(\ref{integrals}), it is possible to express $x(t)$,
$y(t)$, and $z(t)$ in the form of an integral as in BL06. Nevertheless the energy
conservation only gives an implicit relation between those variables and
the computation remains tedious.
For this reason, we give here an algorithm that enables to compute the
two frequencies in a simple way using the numerical integration
of the secular equations (\ref{eq.sec}).
The method leads to an arbitrary high precision since it necessitates
the integration over one nutation period only. 

We assume that at $t_0=0$, the initial volume $v$ (\ref{eq.xyz}) is not zero, and let 
$x$ (for example) be the variable with the largest variation rate, $\dot{x}(t_0)$.
Using the method of H\'enon (1963), we search for the first time $t> t_0$ when 
$(x(t), \dot x (t))= (x(t_0), \dot x (t_0)) $.
We integrate the system (\ref{eq.sec})
until  
\be
\left\{\EQM{
x_{n-1} &<& x_0 \cr
x_n &\ge& x_0 \cr
}\right.
\quad \text{ if }\dot x(t_0)>0
\ee
or
\be 
\left\{\EQM{
x_{n-1} &>& x_0 \cr
x_n &\le& x_0 \cr
}\right.
\quad \text{ if }\dot x(t_0)<0\ .
\ee 
We then   change the time variable to $x$ and 
integrate
\be
\EQM{
\frac{dt}{dx} &=& \frac{1}{\dot x(x,y,z)}, \crm
\frac{dy}{dx} &=& \frac{\dot y(x,y,z)}{\dot x(x,y,z)}, \crm
\frac{dz}{dx} &=& \frac{\dot z(x,y,z)}{\dot x(x,y,z)}, \crm
\frac{d\theta}{dx} &=& \frac{\epsilon\sqrt{\Theta(x,y,z)}}{\dot x(x,y,z)}\
,
}
\ee
from $x_n$ to $x_0$. The latter equation comes from (\ref{eq.Theta}) and 
will provide the rotation angle of the vectors over one nutation period 
(knowing the initial angle $\theta(t_0)$). We thus have the nutation
period $t=T$ and $\theta_T=\theta(T)-\theta(t_0)$. The precession period
is simply given by
\be
T'=\frac{2\pi}{\theta_T}T\ .
\ee

\section{Analytical approximation}
\llabel{sec:anap}
In this section we give an analytical approximation of the secular
evolution. So far, only general features of the solutions have been
obtained. Here analytical approximations of the two
frequencies that appear in the problem as well as their amplitudes 
are computed. The two frequencies being the global precession and the
nutation.

In an invariant frame where the third axis is aligned with the direction
$\w_0$ of the total angular momentum, we can write
\be
\w = \bpm \xi \\ \eta \\ \zeta \epm, \qquad
\w_1 = \bpm \xi_1 \\ \eta_1 \\ \zeta_1 \epm, \qquad
\w_2 = \bpm \xi_2 \\ \eta_2 \\ \zeta_2 \epm
\ee
where
\begin{multline}
\zeta   = \Frac{\gamma  +\beta x+\alpha y}{W_0}, \quad
\zeta_1 = \Frac{\gamma x+\beta  +\alpha z}{W_0}, \quad \\
\zeta_2 = \Frac{\gamma y+\beta z+\alpha  }{W_0}\ .
\end{multline}
The evolution of the projections on the complex plane orthogonal to $\w_0$
\be
\mz   = \xi + i\eta, \quad
\mz_1 = \xi_1 + i\eta_1, \quad
\mz_2 = \xi_2 + i\eta_2,
\llabel{eq:mz}
\ee
is obtained from the secular equations (\ref{eq.sec}), and yields to
\be
\frac{d}{dt}\begin{pmatrix} \mz \\ \mz_1 \\ \mz_2 \end{pmatrix} =
i M\begin{pmatrix} \mz \\ \mz_1 \\ \mz_2 \end{pmatrix}
\llabel{eq.proj}
\ee
where
\be
M = \bpm
-\frac{p}{\gamma}\zeta_1-\frac{q}{\gamma}\zeta_2 &
 \frac{p}{\gamma}\zeta         &
 \frac{q}{\gamma}\zeta        \\
 \frac{p}{\beta} \zeta_1       &
-\frac{p}{\beta} \zeta  -\frac{s}{\beta}\zeta_2 &
 \frac{s}{\beta} \zeta_1      \\
 \frac{q}{\alpha}\zeta_2       &
 \frac{s}{\alpha}\zeta_2       &
-\frac{q}{\alpha}\zeta  -\frac{s}{\alpha}\zeta_1
\epm
\ee
and $(p, q, s)$ are defined in (\ref{eq.pqs}). $M$ is a real
matrix with periodic coefficients. As it is not possible to obtain a simple
analytical solution of this system, we make a crude approximation.
Hereafter we replace the matrix $M$ by the constant matrix $\tM$
obtained by substituting $(x, y, z)$ by their average
\be
\tM = M(\tilde{x}, \tilde{y}, \tilde{z})\ .
\llabel{eq.meanM}
\ee
The solution of (\ref{eq.proj}) is thus straightforward. 
It is easy to verify that $(\zeta, \zeta_1, \zeta_2)$ is an eigenvector
of $\tM$ with eigenvalue 0. The other eigenvalues are then the solutions of 
\be
\lambda^2 - {\bf T} \lambda + {\bf P} = 0
\ee
where ${\bf T}$ is the trace of $\tM$ and 
\be
{\bf P} = \left(
  \frac{\zeta  }{\alpha\beta}
+ \frac{\zeta_1}{\gamma\alpha}
+ \frac{\zeta_2}{\beta\gamma}
\right)
(pq\zeta + sp\zeta_1 + qs\zeta_2) \ .
\ee
Let $\Omega$ and $\Omega+\nu$ be the other two eigenvalues such that
\be
\Omega = \frac{{\bf T}+\sqrt{{\bf T}^2-4{\bf P}}}{2}, \qquad
\nu = -\sqrt{{\bf T}^2-4{\bf P}} \ .
\label{eq.omeganu}
\ee
The system possesses three eigenmodes
\be
\fu e^{i\psi}, \qquad \mr e^{i(\Omega t+\Phi)}, \qquad 
\ms e^{i[(\Omega+\nu)t+\Phi+\phi]},
\ee
with eigenvectors
\be
e_0 = \bpm \zeta \\ \zeta_1 \\ \zeta_2 \epm, \qquad
e_1 = \bpm 1 \\ \lambda \\ \mu \epm, \qquad
e_2 = \bpm 1 \\ \lambda' \\ \mu' \epm,
\ee
where $\lambda$, $\lambda'$, $\mu$ and $\mu'$ are real numbers. The
solutions are then
\be
\EQM{
\mz   = \zeta\fu\e^{i\psi}+\e^{i(\Omega t+\Phi)} \Big(\mr+\ms\e^{i(\nu t+\phi)}\Big), \crm
\mz_1 = \zeta_1\fu\e^{i\psi}+\e^{i(\Omega t+\Phi)}\big(\lambda\mr+\lambda'\ms\e^{i(\nu t+\phi)}\big), \crm
\mz_2 = \zeta_2\fu\e^{i\psi}+\e^{i(\Omega t+\Phi)}\big(\mu\mr+\mu'\ms\e^{i(\nu t+\phi)}\big). \crm
\llabel{eq:secmz}
}
\ee
Moreover, $\gamma\mz+\beta\mz_1+\alpha\mz_2=0$ as it is the projection
of $\bW_0$ on a plane orthogonal to $\bW_0$. This implies that the
constant term $(\gamma\zeta+\beta\zeta_1+\alpha\zeta_2)\fu \e^{i\psi}$ is
also null. As $\gamma\zeta+\beta\zeta_1+\alpha\zeta_2=W_0$, we have
necessarily $\fu=0$. The solutions are thus
\be
\EQM{
\mz   &=& \e^{i(\Omega t+\Phi)}\Big(\mr+\ms\e^{i(\nu t+\phi)}\Big), \crm
\mz_1 &=& \e^{i(\Omega t+\Phi)}\Big(\lambda\mr+\lambda'\ms\e^{i(\nu t+\phi)}\Big), \crm
\mz_2 &=& \e^{i(\Omega t+\Phi)}\Big(\mu\mr+\mu'\ms\e^{i(\nu t+\phi)}\Big). \crm
}
\llabel{eq.sol}
\ee
In this approximation, the three axes $(\w, \w_1, \w_2)$ describe
circular motions with nutation frequency $\nu$ around the three poles 
$(\cP, \cP_1, \cP_2)$ that precess uniformly with precession frequency
$\Omega$ around the total angular momentum $\bW_0$. As it was previously
said, the three poles $(\cP,\cP_1,\cP_2)$ remain always coplanar with
$\bW_0$.

\subsection{Initial conditions}
The preceding section shows that the solutions (\ref{eq.sol}) depend
only on four real numbers $\mr$, $\ms$, $\Phi$ and $\phi$. At the origin
of time $(t=0)$ we can choose two vectors, for instance
\be
\mz_0 = \e^{i\Phi}\Big(\mr+\ms\e^{i\phi}\Big),  \quad\text{and}\quad
\mz_{10} = \e^{i\Phi}\Big(\lambda\mr+\lambda'\ms\e^{i\phi}\big)
\ee
from which we derive
\be
\mr\e^{i\Phi} = \frac{\lambda'\mz_0-\mz_{10}}{\lambda'-\lambda},
\quad{\text{and}}\quad
\ms\e^{i\phi} = \frac{\lambda\mz_0-\mz_{10}}{\lambda-\lambda'}.
\label{eq.ini}
\ee
The computation of $\lambda$ and $\lambda'$ requires the knowledge of the
averaged values $\tilde{x}$, $\tilde{y}$ and $\tilde{z}$, but it can
easily be done by iteration, starting with the initial values, that is,
for the first iteration
\be
\tilde{x}=x(t=0), \qquad
\tilde{y}=y(t=0), \qquad
\tilde{z}=z(t=0).
\ee
In our computations, we found that one iteration after this first try
with the initial conditions was sufficient to obtain a satisfactory
approximation for the frequency amplitudes and phases of the solution
(see Tables \ref{Tabf}, \ref{Tabc}, \ref{Tabd}).

\subsection{Second order expansion}
The whole previous study has been made with an Hamiltonian expanded 
up to the fourth degree in $R/r$ (\ref{eq.HT}), (\ref{eq.HE}), (\ref{eq.HIa}), (\ref{eq.HIb})
and (\ref{eq.HIc})
\be
\Ham = \Ht + \He + \Hi^{(0)} + \Hi^{(2)} + \Hi^{(4)}.
\ee
When the body-body interactions are neglected,
we can restrict the analysis to the second degree in $R/r$. The secular 
Hamiltonian then simplifies to
\be
\tHs = -\frac{\tfa}{2}x^2-\frac{\tfb}{2}y^2 + \const
\ee
where
\be
\tfa = k_3m_2\fCa \quad{\text{and}}\quad 
\tfb = k_3m_1\fCb
\ee
and with 
\be
\EQM{
k_3 = \frac{3}{2}\frac{\cG}{a^3(1-e^2)^{3/2}} \crm
\fCa = \left(1-\frac{3}{2}\sin^2J_1\right)\left(C_1-\frac{A_1+B_1}{2}\right) \crm
\fCb = \left(1-\frac{3}{2}\sin^2J_2\right)\left(C_2-\frac{A_2+B_2}{2}\right).
}
\ee
The secular equations (\ref{eq.sec}) become
\be
\EQM{
\dot\w = -\frac{\tfa x}{\gamma} \w_1 \times \w 
         -\frac{\tfb y}{\gamma} \w_2 \times \w \ , \crm
\dot\w_1 = -\frac{\tfa x}{\beta} \w \times \w_1 \ , \crm
\dot\w_2 = -\frac{\tfb y}{\alpha} \w \times \w_2 \ ,
}
\ee
where $\gamma$, $\beta$ and $\alpha$ are still the angular momentum of
the orbit, of the rotation of the primary and of the rotation of the secondary
respectively. In that case, the matrix $M$ giving the evolution of the
projection of the three vectors $\mz$, $\mz_1$ and $\mz_2$ becomes
\be
\uM =\bpm
-\frac{\tfa x}{\gamma}\zeta_1-\frac{\tfb y}{\gamma}\zeta_2 & \frac{\tfa x}{\gamma}\zeta
& \frac{\tfb y }{\gamma}\zeta \\
\frac{\tfa x }{\beta}\zeta_1 & -\frac{\tfa x }{\beta}\zeta & 0 \\
\frac{\tfb y }{\alpha}\zeta_2 & 0 & -\frac{\tfb y }{\alpha}\zeta 
\epm.
\ee
Now we use the same trick as in the equation (\ref{eq.meanM}), that is
we replace the matrix $\uM$ by the constant matrix $\tilde{\uM}$
\be
\tilde{\uM} = \uM(\tilde{x},\tilde{y},\tilde{z})
\ee
where $(x,y,z)$ have been substituted by their average.
The vector $\trans{(\zeta,\zeta_1,\zeta_2)}$ is still an eigenvector for
the eigenvalue 0. The characteristic equation is now 
\be
\lambda^2-\tbT\lambda+\tbP = 0
\ee
where 
\be
\EQM{
\tbT = -\frac{\tfa x}{\gamma}\zeta_1 - \frac{\tfb y}{\gamma}\zeta_2
-\frac{\tfa x}{\beta}\zeta - \frac{\tfb y}{\alpha}\zeta \crm
\tbP = \tfa \tfb xy \zeta\left(
\frac{\zeta}{\alpha\beta} + \frac{\zeta_1}{\gamma\alpha} + \frac{\zeta_2}{\beta\gamma}
\right).
}
\ee
These expressions give simpler formulas for the frequencies, although
they still have the same form
\be
{\Omega} = \frac{\tbT+\sqrt{\tbT^2-4\tbP}}{2}, \qquad
{\nu} = -\sqrt{\tbT^2-4\tbP} \ .
\ee

\section{Global precession of a $n$-body system}
\llabel{sec.back}

We have seen that the secular motion of a two  solid body system  can, as  in BL06, be decomposed 
in a uniform precession of  angular motion $\Omega$, and a periodic  motion of frequency $\nu$.
In fact, this can be extended to a very general system of $n$ solid bodies in gravitational interaction.
The following result, which is of very broad application, 
is a consequence of the general angular momentum reduction  in case of regular, quasiperiodic, 
motion.

\begin{prop}
\llabel{propc}
  Let ${\cal S}$ be a system of $n+1$   bodies of mass $m_i, (i=0,\dots n)$ 
in gravitational interaction, with $n_s$ solid bodies 
among them ($n_s \leq n+1$). Then, in a reference frame centered on one of  the  bodies, 
and for a regular quasiperiodic solution of ${\cal S}$,
there exist a constant precession rate $\Omega$, such that 
any vector  $Z \in \{\rbl_i, \tr_i, \bI_j,\bJ_j,\bK_j, \bG_j; i=1,\dots n; j=1,\dots n_s \}$ has a temporal evolution that can be decomposed  as
\be
Z(t) = \cR_3(\Omega t) \tilde Z^{(\nu)}(t) \ ,
\ee
where $\cR_3(\Omega t)$ is a uniform precession around the total angular momentum $\bW_0$ with constant rate $\Omega$,
and where $\tilde Z^{(\nu)}(t)$ can be expressed in term of quasiperiodic series of $3(n+n_s)-2$  frequencies  $(\nu_k)$. 
We will call $\Omega$ the global precession rate of the system ${\cal S}$. 
\label{prop3}
\end{prop}

\begin{proof}
Let us consider a  general system of $n+1$ bodies of mass $m_i, (i=0,\dots n)$ in gravitational interaction, with $n_s$ solid bodies 
among them ($n_s \leq n+1$). 
This is a $3(n+1+n_s)$ degree of freedom (DOF) system. Due to the translation invariance of the 
system, it can be reduced to $N=3(n+n_s)$ DOF using the 
coordinates  centered on one of the bodies (the one of mass $m_0$ for example). This 
heliocentric reduction can be made in canonical form, preserving the Hamiltonian structure
of the equations (see Laskar and Robutel, 1995).

The full Hamiltonian of the system, as expressed in (\ref{eq.Hfull}) 
is then a function of  the vectors $(\rbl_i, \tr_i, \bI_j,\bJ_j,\bK_j, \bG_j), i=1,\dots n; j=1,\dots n_s,$
that depends uniquely of the scalar products of theses vectors. 
Moreover, the total angular momentum $\bW_0$ (\ref{integrals}) is conserved. 

This system, as for the usual reduction of the node, can be reduced to a system of $N-2$ degrees of freedom.
A first reduction  to $N-1$ DOF can be achieved by  using a reference frame $(\bi,\bj,\bk)$ such that 
$\bk$ is collinear with $\bW_0$ and $\dop{\bk}{\bW_0}$ is positive. This  partial reduction 
is based uniquely on the fixed direction of the angular momentum (Malige \etal, 2002).
With this reference frame, all quasiperiodic solutions of the system can   be expressed in 
term of only $N-1$ fundamental frequencies.

In this fixed $(\bi,\bj,\bk)$ reference  frame, we can use canonical coordinates that 
are well adapted for both the orbital and rotational motions. 
Namely, we shall use the Andoyer coordinates for the solid bodies $(L,G,H,l,g,h)$
(Fig. \ref{Figb}), and the equivalent Delaunay coordinates for the orbital motions 
$(\Lambda=\beta\sqrt{\mu a},\Gamma=\Lambda\sqrt{1-e^2},\Theta=\Gamma\cos i, M, \omega, \theta)$ where
$(a,e,i,M,\omega, \theta)$ are the usual elliptical elements (semi-major axis, eccentricity, inclination 
of the orbit with respect to the  $(\bi,\bj)$ plane, mean anomaly, argument of periapse, longitude of the 
ascending node). For any given body of mass $m_i, i\neq 0$ ,
$\beta_i = m_0m_i/(m_0+m_i)$ is the reduced mass, and $\mu_i= G(m_0+m_i)$ the related gravitational constant.
For any $X_i \in \{\rbl_i, \tr_i; i=1,\dots n \}$,   or $Y_j \in \{\bI_j,\bJ_j,\bK_j, \bG_j; j=1,\dots n_s\}$,  one can then 
write 

\be
\EQM{
X_i & = \cR_3(\theta_i) X_i'(\Lambda_i, \Gamma_i, \Theta_i, M_i, \omega_i) \ ;\cr 
Y_j &=  \cR_3(h_j) Y'_j(L_j,G_j,H_j,l_j,g_j) \ .
}
\ee

Let us now select one angle among the $\theta_i,h_j$ ($\theta_1$ for example) and perform the usual
symplectic linear change of variable 
\be
\EQM{
\theta'_1&=\theta_1           \ ;  \qquad  &\Theta_1' &= \sum_i \Theta_i + \sum_j H_j \crm
\theta'_i&=\theta_i-\theta_1  \ ;  \qquad  &\Theta_i' &=\Theta_i  \ \hbox{for } i\neq 1 \crm
h_j' &= h_j -\theta_1   \ ;  \qquad  &H'_j&= H_j 
}
\label{eq.varp}
\ee

As the Hamiltonian (\ref{eq.Hfull}) depends only on the scalar products of $X_i$ and $Y_j$, it can be as well 
expressed in term of scalar products of 
\be
\tilde X_i = \cR_3(-\theta_1) X_i  \ ; \qquad   \tilde Y_j = \cR_3(-\theta_1) Y_j \ .  
\ee

Expressed in term  of the new variables (\ref{eq.varp}), one can see that the coordinate $\theta'_1$ is now 
ignorable with an associated constant action being the 
modulus of the total angular momentum ($\Theta'_1 = \norm{\bW_0}$). The number of 
DOF of the  system, expressed in the new coordinates 
($\Lambda_i, \Gamma_i, \Theta_i', M_i, \omega_i,\theta_i', L_j,G_j,H_j',l_j,g_j,h'_j$) is now $N-2$, with 
one constant parameter, $\Theta'_1$.
Let us now consider a quasiperiodic solution of the above   $N-2$ DOF  system. All vectors $\tilde X_i, \tilde Y_j$
will be expressed in term of quasiperiodic functions on $N-2$ independent frequencies $\nu_k, (k=1,\dots N-2)$. Finally, 
 $\theta'_1$ evolution   is given by
 
 \be
  {d\theta'_1\over dt}= \Dron{H}{\Theta'_1}(\Lambda_i, \Gamma_i, \Theta_i', M_i, \omega_i,\theta'_{i, i\neq 1}, L_j,G_j,H_j',l_j,g_j,h'_j) \ .
 \ee

Thus $ \dot \theta'_1 (t)$ is also a quasiperiodic expression depending on the $N-2$ frequencies $\nu_k$.

\be
 {d\theta'_1 \over dt} = \sum_{(k)} \alpha_{(k)} \exp (i < k,\nu> t) \ ,
\ee
where $(k)$ is a ($N-2$) multi index. Let $\Omega = \alpha_{(0)}$ be the constant term of this series. We have then 
\be
 {d\theta'_1 \over dt} = \Omega + \sum_{(k)\neq (0)} \alpha_{(k)} \exp (i < k,\nu> t) \ ,
\ee
and thus
\be
\theta'_1(t) = \Omega t + f_{(\nu)}(t) \ ,
\ee
where $f_{(\nu)}(t)$ is a $(N-2)-$periodic function with frequencies $(\nu_k)$. The original vectors $X_i, Y_j$ can then be 
expressed as 
\be
\EQM{
 X_i &= \cR_3(\theta_1) \tilde X_i  &= \cR_3(\Omega t) \cR_3(f_{(\nu)}(t) )\tilde X_i    &=   \cR_3(\Omega t) \tilde X_i^{(\nu)} \ ,\cr
 Y_j &= \cR_3(\theta_1) \tilde Y_j  &= \cR_3(\Omega t) \cR_3(f_{(\nu)}(t) )\tilde Y_j    &=   \cR_3(\Omega t) \tilde Y_j^{(\nu)} \ ,
 }
\ee
where $ \tilde X_i^{(\nu)}, \tilde Y_j^{(\nu)}$ can be expressed in term of $(N-2)-$periodic function with frequencies $(\nu_k)$. 
This ends the proof of the proposition.
\end{proof}

\begin{consequence}
A consequence of this  result is that for a quasiperiodic solution of the general two body 
problem that we are considering here ($n=1, n_s=2$), 
 the components of any vectors $\rbl, \tr, \bI_j,\bJ_j,\bK_j, \bG_j$, should express 
 as  quasiperiodic functions of the precessing frequency  $\Omega$ and of 7 frequencies $\nu_k, k=1,\dots 7$, the precession frequency $\Omega$ 
 appearing in all terms with coefficient 1. This is actually what is observed on some examples in the next section (Tables \ref{Tabba} and \ref{Tabbb}).
 One should note that the same results hold for the three body problem studied in BL06 (with  $n=2, n_s=1$).
\end{consequence}

It is also useful to remark that the value of $\Omega$ is independent of the $\nu_k$, i.e.
any commensurable relation between  $\Omega$  and the $\nu_k$ has no effect on the dynamics of the system,
in the sense that it will not affect the regularity of the solutions. On the other hand, 
in the  case of a single $\nu_k$ frequency (as for the secular system), a rational ratio $\Omega/\nu$ 
will lead to a periodic solution in the fixed reference frame $(\bi, \bj, \bk)$. 
We prefer here to speak of geometric resonance instead of dynamical resonance,
as there is no coupling between the two degrees of freedom of frequency $\Omega$ and $\nu$.

\section{Application}
\llabel{sec.res}
In this section we compare our rigorous results on the averaged system 
and our analytical approximations of the solutions of the same system with the
integration of the full Hamiltonian (\ref{eq.HT}), (\ref{eq.HE}),
(\ref{eq.HIa}), (\ref{eq.HIb}) and (\ref{eq.HIc}) on two different
binary systems $I$ and $I\!I$ (see table~\ref{Taba} and \ref{Tabab}). The
physical and orbital parameters of the system $I\!I$ are those of the binary
asteroid 1999 KW4 studied in FS08. We choose this system in order to
compare our results with FS08. In this case, the rotation of the
satellite is taken to be synchronous. As our analytical results were obtained
assuming the satellite rotation asynchronous, we create a system $I$
from the system $I\!I$ where the rotation of the secondary has been
sped up by a factor 3. Since the orbit is circular and the initial
rotation axes aligned with the axes of maximum inertia, the system $I\!I$ 
is highly degenerated. To get a more general system where all the
fundamental frequencies will actually exist, we changed the initial
Andoyer angles and the eccentricity. But then, because of its strong
triaxiality, the evolution of the satellite orientation becomes chaotic
(Wisdom, 1987).
As here, we are concerned only with on regular behaviors, we thus
decreased the satellite 
triaxiality and increased the semi-major axis in order to obtain a
generic example of regular solution.


\subsection{Numerical experiments}
\subsubsection{Frequency analysis}
The quasiperiodic decomposition of our numerical integrations was
obtained using the frequency analysis developed by Laskar (Laskar, 1988,
2005). As our systems contain a large range of frequencies
going from 0.07 rad$\cdot$day$^{-1}$ to 109 rad$\cdot$day$^{-1}$, we decided to run
twice each integration with two different output time steps $h=0.1$ days
and $h'=0.1001$ days. These two time steps do not fulfilled the Nyquist condition
for the largest frequency. Nevertheless, it is possible to recover the true 
value $\nu_0$ of the frequency using the following trick (Laskar, 2005). For a real x, let
denote $[x]$ the real such that
\be
-\pi < [x] \le \pi.
\ee
Let $\nu$ and $\nu'$ be respectively the frequencies measured on the 
integration with the time step $h$ and $h'$. The true frequency is
given by
\be
\nu_0 = \nu + \frac{[k]}{h}
\ee
where
\be
k = \frac{h}{h'-h}((\nu'-\nu)h'-[\nu'h'-\nu h]).
\ee

\prep{\taba}
\prep{\tabab}
\prep{\tabe}
\prep{\tabf}
\prep{\tabba}
\prep{\tabbb}
\prep{\tabc}
\prep{\tabd}

\subsubsection{System I -- doubly asynchronous case}
\paragraph{Full Hamiltonian}
We integrated the system $I$ over a time span of $2\,000$ days and performed a 
frequency analysis as described above. This system contains a priori 9
degrees of freedom. Three coordinates for the orientation of each body and three
coordinates for the orbit. But there is a relation between all these
coordinates given by the conservation of the total angular momentum.
There are thus only 8 degrees of freedom. Hence the system contains 8
fundamental frequencies (cf table~\ref{Tabe}).

These frequencies can be divided into four main categories: $1)$ the
secular frequencies containing the precession $\Omega$ and the nutation $\nu$;
 $2)$ the orbital frequencies with the periapse precession rate $\hat{\omega}$ and 
the mean motion $n$; $3)$ 
the frequencies of the primary $\hat{g}_1$ and $\hat{l}_1$ associated respectively 
to the Andoyer angles $g_1$ and $l_1$; $4)$ the same frequencies for the
secondary $\hat{g}_2$ and $\hat{l}_2$.

Table~\ref{Tabba} displays the frequency decomposition in the form 
$\sum A_j\exp{i(\nu_j t+\varphi_j)}$ of the motion of
$\mz$, $\mz_1$ and $\mz_2$ (\ref{eq:mz}), the projections of $\w$, $\w_1$ and
$\w_2$ on the complex plane orthogonal to $\bW_0$. The second column
shows that all the frequencies are combinations of the 8 fundamental
frequencies.

Moreover, we verify our proposition saying that in a frame rotating uniformly 
with the precession rate $\Omega$, the system loses one degree of freedom, see 
section \ref{sec.back}. Indeed, the frequency $\Omega$ appears in all
the terms with the same order 1.


\paragraph{Averaged Hamiltonian}
In the frequency decomposition of the motion of $\w_1$ in Table~\ref{Tabba}, 
the nutation is only the $4^{\rm th}$ term. To check the validity of the
averagings, we integrated the averaged Hamiltonian (\ref{eq.Hs}) on the
same time span (2\,000 days) and we performed the same frequency
analysis. Initial rotation rates, semi-major axis and eccentricity are
average values computed on the numerical output of the full integration.
Initial inclination, obliquities and ascending nodes were obtained from
the amplitudes and the phases of the frequency analysis in
Table~\ref{Tabba}.

\prep{\fige}

Table~\ref{Tabc} displays the comparison between the frequency
decomposition of the output of the averaged Hamiltonian and of the full
Hamiltonian (columns 3--8). For the comparison, only the secular terms
were extracted from the analysis of the full integration. The second
column confirms our analytical result saying that the averaged motion
contains only 2 fundamental frequencies: the precession $\Omega$ and the
nutation $\nu$; and that in a frame rotating with the precession frequency,
only the nutation remains. The columns 4,5 and 7,8 show the strong
agreement between the secular approach and the full integration. Even
low amplitude terms such as $\Omega+2\nu$, albeit at the $57^{\rm th}$
position in the decomposition of $\w$ in the full integration, are
recovered with good amplitude and phase in the regular system.

The two last columns of table~\ref{Tabc} give the complex amplitudes of
the secular motion obtained with the analytical approximation of
section~\ref{sec:anap}. As in this approximation, the nutation is assumed
to be a uniform rotation, there are only two terms in the description of
the secular motion. Nevertheless, we see that this approximation is also
in good agreement with the integration of the full Hamiltonian and of
the averaged Hamiltonian.

In Table~\ref{Tabf} are given the values of the secular
frequencies for systems $I$ and $I\!I$, obtained either from the
integration of the full Hamiltonian, from the integration of the
averaged Hamiltonian, or with the analytical approximations (\ref{eq.omeganu}).
The precession rates are in agreement
within 0.3\% and the nutation frequencies within 5\%.

Figure~\ref{Fige} represents the trajectories of the unit vectors on
the plane orthogonal to the total angular momentum $\bW_0$. In the left
panel, the frame is fixed, it thus corresponds to ${\cal W}(t)$, see
section~\ref{sec.global}. 
We see that the evolutions of $\w$ in red and
$\w_1$ in green are dominated by the precession: their orbits are
quasi-circular, whereas the orbit of $\w_2$ in blue contains a large
nutation as it can be checked in the frequency analysis table~\ref{Tabba}.
The right panel shows the same orbits but in a frame rotating with the
precession rate $\Omega$, it corresponds to $\tilde{\cal W}(t)$.
It emphasizes the nutation loops. Zooms on the nutation of the orbit and
of the primary are plotted on the furthest right. The solid curves
are the output of the averaged Hamiltonian. The analytical
approximations cannot be distinguished from the averaged output.
These averaged solutions are good
approximations of the motion of $\w$ and $\w_2$ but the agreement
does not seem to be as good for $\w_1$.
Indeed, table~\ref{Tabba} shows that high frequencies have larger
amplitudes than the secular nutation.

Because of this amplitude issue for $\w_1$ in Fig.~\ref{Fige}, we
decided to filter our full integration with a low-pass filter to see if we
could get back the averaged integration. In this scope, we reintegrated
the full Hamiltonian over a time span of 20 days with an output
time step of 30 min. We then filtered the output with a cutoff frequency
equal to 4 rad/day. The filtered trajectories are displayed in
Fig.~\ref{Figf}. The nutation amplitude of $\w_1$ is now well
retrieved. After a small change in the initial conditions that corresponds to a 
decrease of only $3.6\arcsec$ of the initial obliquity of the
primary in the averaged Hamiltonian, we get back the filtered full
Hamiltonian (see Fig.~\ref{Figg}).

\subsubsection{System II -- asynchronous-synchronous case}
For this second experiment,
we took the same initial conditions as FS08 (table~\ref{Tabbb}).
The primary has an asynchronous rotation whereas the secondary rotates
synchronusly. The
difference between our study and FS08 is that we expanded the
Hamiltonian up to the fourth order in $R/r$ where $R$ is the radius of
one body and $r$ the distance between them. We performed the same 
frequency analysis as with system $I$. We get also 8 fundamental
frequencies. Because the resonance, the frequencies associated to the
secondary are not $\hat{g}_2$ and $\hat{l}_2$ anymore since they are
in that case combinations of the other 6 fundamental frequencies.
The two new frequencies correspond to the horizontal and vertical
libration of the secondary: $\hat{\psi}_2$ and $\hat{\theta}_2$
respectively.

Table~\ref{Tabbb} presents the frequency analysis performed on this
system. The result of section {\ref{sec.back}} is still valid,
in a
frame rotating with the precession rate, the system loses one degree of
freedom. We confirm that this result does not depend on the 
resonances in the reduced problem.

The averaged Hamiltonian and the analytical approximation were not
specifically
written for such a resonant case. Regardless of this fact, the results of
the averaged Hamiltonian and of the analytical approximation applied to
this system are summarized in table~\ref{Tabd}. It is remarkable that
the first two amplitudes of each vector are in good agreement with the
full integration. Nevertheless, the third amplitude of $\w_2$ is wrong
by a factor 3. The values of these secular frequencies are given at the
bottom of table~\ref{Tabf}. The use of the averaged Hamiltonian or of the
analytical approximations leads to an error on the precession rate equal
to 1\% and on the nutation rate equal to 24\%.

\prep{\figf}
\prep{\figg}

\subsection{Comparison with FS08}
\subsubsection{Numerical results}
In FS08, Fahnestock and Scheeres expanded the Hamiltonian up to the second order
in $R/r$. They find that motions of binary asteroids such 
as 1999 KW4 are combinations of four modes with their
respective fundamental frequency. The first and fastest mode corresponds to 
the rotation of the primary around its axis. The second mode coincides
with the orbital motion which has the same period as the rotation of the
secondary around its axis. The third mode is said to be an excitation of the
satellite's free precession dynamics and has a period of $\approx 188$ h. 
The corresponding frequency would be $\approx 0.802$ rad/day.
The last mode is identified as the precession motion.

Our results generally agree with the analysis of 
Fahnestock and  Scheeres. Nevertheless, several 
frequencies are missing in their analysis, probably 
because of the degeneracy of their initial conditions. As the initial
eccentricity is close to 0, and the angular momenta along the axes of
maximum inertia, the first terms in the frequency decompositions are
combinations of $\hat{\omega}+n$ which corresponds to their orbital
frequency, and of $\hat{g}_1+\hat{l}_1$ which corresponds to their rotation of
the primary, see table~\ref{Tabbb}. On the other hand, we do not find
their third mode of frequency $\approx0.802$ rad/day.

\subsubsection{Solid-point interaction}
Fahnestock and Scheeres found also that the spin axis of the primary 
and the orbital plane precess at the same rate.
They derived an analytical expression for this precession rate, see
their equation (76). Their result corresponds in fact to the solution
of the single planet case that is already described in BL06 and which
does not require the more elaborated formalism developed here.
Indeed, as they expanded the potential up to the second order only, they
canceled the effect of the orientation of the secondary on the
precession of the primary $(\fc=\fd=\fe=\ff=\fg=0)$. Moreover, as they fixed the 
orientation of the secondary with the orbit, the secondary does not 
influence the orbit ($y=(\dop{\w}{\w_2})=1$).
We recall here the derivation of this frequency as given in BL06.
With the assumption of a point mass satellite, the Hamiltonian becomes
\be
\Ham = -\frac{\tfa}{2}x^2,
\ee
with $x=(\dop{\w}{\w_1})$ and
\be
\EQM{
\dot{\w} = -\frac{\tfa}{\gamma}(\dop{\w}{\w_1})\w_1\times\w, \crm
\dot{\w_1} = -\frac{\tfa}{\beta}(\dop{\w}{}\w_1)\w\times\w_1.
}
\ee
This reduced problem has 5 independent integrals given by
\be
\EQM{
\norm{\w}  = 1 \crm
\norm{\w_1}  = 1 \crm
\gamma \w + \beta \w_1 = \bW_0 \ .
}
\ee
As $ x=\w\cdot\w_1$ is constant, the system is
trivially integrable. We have indeed
\be
\dot\w    =  \Omega_0 \w_0 \times \w  \ , \quad
\dot\w_1  = \Omega_0 \w_0  \times \w_1   \ ;
\ee
where $\w_0=\bW_0/ \norm{\bW_0}$ is the unit vector in the direction of
the total angular momentum $\bW_0$, and
\be
\Omega_0 = -\Frac{\tfa x}{\A}\sqrt{1+\Frac{\A^2}{\B^2} + 2 \Frac{\A}{\B}
x } \ .
\llabel{eq.Om0}
\ee

Both vectors $\w, \w_1$ thus precess uniformly around the total angular
momentum direction $\w_0$ with constant precession rate $\Omega_0$. The
correspondence with the notations of the equation (76) of FS08 is
\be
\EQM{
\frac{\tfa}{\gamma} = \frac{3}{2}\frac{\sqrt{\mu}}{a^{7/2}(1-e^2)^2}
(I_{1}-I_{\rm eq})\left(1-\frac{3}{2}\sin^2J_1\right) \crm
x = \cos(\delta+i) \crm
\sqrt{1+\Frac{\A^2}{\B^2} + 2 \Frac{\A}{\B}x } =
\frac{\sin(\delta+i)}{\sin i}.
}
\ee
 
{\bf Remark.} The factor $(1-(3/2)\sin^2J_1)$ is not in FS08 because
FS08
assumes that the primary angular momentum is aligned with its figure
axis $(J_1=0)$. This is not the case in this paper where we do not
require this simplification.

\section{Conclusions}

We have shown here that  the general framework developed in BL06 
applies as well to the
problem of two rigid bodies orbiting each other. This formalism enables us
to obtain the long term evolution of the spin axis of the two bodies as
well as the evolution of the orientation of the orbital plane.
The two bodies can be very general, with strong  triaxiality, 
and their rotation vector is 
not necessary aligned with their axis of maximum inertia. 
The gravitational potential is expanded up to the fourth order so as to
keep the direct interaction between the orientation of the two bodies, and 
as in BL06, the evolution
of their spin axis is   obtained after a suitable averaging.

We found that the secular evolution is composed of two periodic motions:
a global precession of the three angular momenta and nutation loops. As
in BL06, the nutation loops are symmetric with respect to a plane
containing the total angular momentum and precessing with the global
precession frequency. We gave analytical approximations of these
frequencies.

We performed a frequency analysis (Laskar, 1988, 2005) on a numerical 
integration of the full Hamiltonian.
We chose the typical binary asteroid system 1999 KW4 already analyzed in FS08.
We retrieved the precession and the nutation motions predicted by the
secular Hamiltonian and estimated by the analytical approximations.
On a non resonant system, derived from 1999 KW4, the secular solution, and 
the  analytical results agree extremely well with the full 
solution. This is still the case to a lesser extent with 
the more specific case of 1999 KW4, which is in 1:1 spin-orbit
resonance.
In a further work, we could consider in a more precise way the possible resonances.
In that case, some of the averagings need  to be done in a 
different way, probably leading to less symmetrical,  more complex, expressions.
The main goal reached by the present paper  was
to search, in this apparently difficult problem of two solid bodies in interaction, 
what was the most simple relevant underlying structure. 
One can now add possible additional  effects, as tidal dissipation, 
and still consider the problem with the present setting.
We thus expect that the results presented here will be helpful for the understanding 
of the general evolution of binary asteroids, or other problems 
of astronomical interest.

In the elaboration of this paper, we came across the very general result given in 
our proposition \ref{propc} which applies to any system of $n$ massive bodies 
(point masses or not) in 
gravitational interaction. This property of the motion states that 
the general regular quasiperiodic motions with $N$ independent frequencies
can be decomposed into a uniform rotation around the total angular momentum,
which we call the global precession,
and in this rotation frame, a quasiperiodic motion with $N-1$ frequencies, 
independent of the global precession frequency.

\subsection*{Acknowledgments}
We thank Franck Marchis for discussions on binary asteroids observations. 
The authors largely benefitted from the interactions and discussions
inside the Astronomy and Dynamical System group at IMCCE.


\subsection*{Appendix A. Gravitational interaction expansion}
The gravitational interaction between the two bodies is given by
(\ref{eq.Hi})
\be
\Hi = -\iint \frac{\cG\,dm_1\,dm_2}{\norm{\rbl+\rbl_2-\rbl_1}}\ .
\ee
The expansion of this potential in Legendre polynomials leads to the
following integrals
\be
\EQM{
H_I^{(0)} &=-\frac{\cG}{r}\iint & {dm_1\,dm_2}\ , \crm
H_I^{(1)} &= 0 \crm
H_I^{(2)} &=-\frac{\cG}{2r^3}\iint & \Big[
   3\rra^2 +3\rrb^2 
   -r_1^2  -r_2^2 \Big]\,dm_1\,dm_2 \crm
H_I^{(3)} &=-\frac{\cG}{2r^4}\iint & \Big[
   5\rra^3      -5\rrb^3      \crm
&& -3r_1^2\rra   +3r_2^2\rrb   \Big]\,dm_1\,dm_2 \crm
H_I^{(4)} &=-\frac{\cG}{8r^5}\iint & \Big[
   35\rra^4        +35\rrb^4                    \crm 
&& -120\rarb\rra\rrb                             \crm
&& -12\rbl_1^2\rarb   +210\rrb^2\rra^2             \crm 
&& -30\rbl_2^2\rra^2  -30r_1^2\rrb^2               \crm
&& -30r_1^2\rra^2   -30r_2^2\rrb^2               \crm
&& +3r_2^4          +3r_1^4          +6r_1^2r_2^2\Big]\,dm_1\,dm_2
}
\ee
where all linear terms in $\rbl_1$ or $\rbl_2$ have been omitted since these
two vectors are expressed relative to the barycenter of the respective
body and their integral vanishes. In the section \ref{sec.potential}, an
additional hypothesis is made on the mass distribution of each body that 
simplifies the potential. They are supposed to be symmetrical relative
to the planes perpendicular to the principal axes of inertia. As a
consequence, the integral of any odd power of $\rbl_1$ or $\rbl_2$ cancels.

\subsection*{Appendix B. Inertia integral}
Inertia integrals of homogeneous ellipsoids are
computed in the following way. Let $a$, $b$, $c$ be the three semi-axes of
a homogeneous ellipsoid $\cal E$ of density $\rho$. The total mass of the ellipsoid
is
\be
m = \frac{4\pi}{3}\rho abc
\ee
and the second order inertia integrals are
\be
\EQM{
\int_{\cal E} \rho x^2\,dx\,dy\,dz = \rho a^3bc\int_{\cB}
X^2\,dX\,dY\,dZ = \frac{1}{5}ma^2\ , \crm
\int_{\cal E} \rho y^2\,dx\,dy\,dz = \rho ab^3c\int_{\cB}
Y^2\,dX\,dY\,dZ = \frac{1}{5}mb^2\ , \crm
\int_{\cal E} \rho z^2\,dx\,dy\,dz = \rho abc^3\int_{\cB}
Z^2\,dX\,dY\,dZ = \frac{1}{5}mc^2\ ,
}
\ee
where $X=x/a$, $Y=y/b$, $Z=z/c$ and $\cal B$ is the unit ball. 
From the definition of the moments of inertia
\be
\EQM{
A &=& \int_{\cal E} (y^2+z^2)\,dm\ , \crm
B &=& \int_{\cal E} (z^2+x^2)\,dm\ , \crm
C &=& \int_{\cal E} (x^2+y^2)\,dm\ ,
}
\ee
we get relations between the semi-axes and the moments of inertia
\be
\EQM{
a^2 &=& \frac{5}{2m}(-A+B+C)\ , \crm
b^2 &=& \frac{5}{2m}( A-B+C)\ , \crm
c^2 &=& \frac{5}{2m}( A+B-C)\ . \crm
}
\llabel{eq.radii}
\ee
General expressions of the inertia integrals are thus
\be
\int_{\cal E} x^iy^jz^k\,dm = \frac{3}{4\pi} m a^ib^jc^k
\int_{\cal B} X^iY^jZ^k\,dX\,dY\,dZ\ ,
\ee
with $a$, $b$, $c$ given by (\ref{eq.radii}).

\subsection*{Appendix C. Averaged quantities}
In this appendix, we give general formulas for the averaging over the
orbital mean motions. The integrals will be computed using the true
anomaly $(\nu)$ as an intermediate variable. We recall first the basic formulas
\be
\EQM{
dM  &=&  \frac{r^2}{a^2\sqrt{1-e^2}} d\nu\crm
\cX &=&  r\cos\nu                        \crm
\cY &=&  r\sin\nu                        \crm
r   &=&  \frac{a(1-e^2)}{1+e\cos\nu}     \crm
}
\ee
where $\cX$ and $\cY$ are the coordinates of a point on a keplerian
orbit in
the  reference frame $(\bi,\bj,\bk)$ with $\bi$ and $\bk$ respectively 
in the direction of periapse and angular momentum. 

\subsubsection*{Intermediate integrals}
In the following, we handle integrals such as Wallis integrals. We
recall their expression. Let
\be
I_n = \frac{1}{2\pi} \int_0^{2\pi} \cos^nt\,dt = 
      \frac{1}{2\pi} \int_0^{2\pi} \sin^nt\,dt
\ee
and 
\be
\EQM{
J_{n,m}&=&\frac{1}{2\pi}\int_0^{2\pi}\sin^mt \cos^nt\,dt \crm
       &=&\frac{1}{2\pi}\int_0^{2\pi}\sin^nt \cos^mt\,dt.
}
\ee
We have then
\be
I_n = \left\{
\EQM{
0 &\quad \text{if $n$ is odd}, \crm
\frac{(2p)!}{2^{2p}(p!)^2} &\quad \text{if $n=2p$, $p\in\mathbb{N}$}\ ,
}
\right.
\ee
and for $p\ge0$, $I_{2(p+1)}$ can be computed using the recurrence formula
\be
I_{2(p+1)} = \frac{2p+1}{2p+2}I_{2p}.
\ee

The integrals $J_{n,m}$ are null whenever $n$ or $m$ is odd, else their
values are a sum of integrals $I_{k}$ 
\be
\EQM{
J_{2p,2q} &=& \frac{1}{2\pi}\int_0^{2\pi}(1-\cos^2t)^p
              \cos^{2q}t\,dt, \crm
          &=& \sum_{k=0}^p (-1)^k \bpm p \\ k \epm I_{2q+2k} \crm
          &=& \sum_{k=0}^q (-1)^k \bpm q \\ k \epm I_{2p+2k}.
}
\ee
The last equality comes from $J_{n,m}=J_{m,n}$.

\subsubsection*{Computation of $\LA1/r^n\RA$ for $n\geq2$}
From these results, we can write
\be
\EQM{
\LA\frac{1}{r^n}\RA&=& \frac{1}{a^n(1-e^2)^{n-3/2}}\frac{1}{2\pi}
\int_0^{2\pi}(1+e\cos\nu)^{n-2}\,d\nu, \crm
&=& \frac{1}{a^n(1-e^2)^{n-3/2}}\sum_{p=0}^{{\rm E}(n/2-1)} 
\bpm n-2 \\ 2p \epm I_{2p}e^{2p}, \crm
&=& \frac{1}{a^n(1-e^2)^{n-3/2}}\sum_{p=0}^{{\rm E}(n/2-1)}
\cA_n(p)e^{2p},
}
\ee
where ${\rm E}(x)$ returns the integer part of $x$ and 
\be
\cA_n(p) = \bpm n-2 \\ 2p \epm\frac{(2p)!}{2^{2p}(p!)^2}.
\ee
The recurrence relation for $\cA_n(p)$ is
\be
\cA_n(p+1) = \frac{(n-2p-2)(n-2p-3)}{(2p+2)^2}\cA_n(p)
\ee
with $\cA_n(0)=1$.

\subsubsection*{Computation of $\LA \cX^m\cY^n/r^l \RA$ for
$l\geq m+n+2$}
In averaging computations we meet integrals in the form
\be
\LA \frac{(\dop{\rbl}{\s_1})^{k_1}\cdots(\dop{\rbl}{\s_j})^{k_j}}{r^l} \RA\
.
\ee
These integrals can be computed from
\be
\EQM{
\LA \frac{\cX^m\cY^n}{r^l} \RA &=& 
\frac{1}{a^h(1-e^2)^{h-3/2}}\frac{1}{2\pi} \crm
&&\hfill\times
\int_0^{2\pi}\cos^m\nu\sin^n\nu(1+e\cos\nu)^{h-2}\,d\nu, \crm
&=& \frac{1}{a^h(1-e^2)^{h-3/2}}\sum_{k=0}^{h-2}
\bpm h-2 \\ 2k \epm J_{n,m+k}e^k,
}
\ee
where $h=l-m-n$ and $J_{n,m}$ defined as previously. This integral is
null whenever $n$ is odd.

\setlength{\leftmargin}{-2.cm}
\setlength{\bibindent}{0.3cm}
\itemindent=-2cm
\parindent=0pt
\setlength{\parindent}{-2cm}
\itemsep=0 pt 
\parsep=0pt
\makeatletter \renewcommand\@biblabel[1]{} \makeatother   
\def\bib#1#2#3#4#5#6#7{\bibitem{#1} {#2}\  {#3}.\ {#4}. {{\it #5}} {{\bf #6,}} {\ #7}.  \par }
\def\bibx#1#2#3#4#5#6#7{\bibitem{#1} {#2}\  {#3}.\ {#4} \  {{\it #5}} {{\bf #6,}} {\ #7}.  \par }
\def\bibB#1#2#3#4#5{\bibitem{#1} {#2}\  {#3}. {{\it #4}.} {(#5, #3)}. \par }
\def\bibll#1#2#3#4#5#6#7{\bibitem{#1} {#2}\  {#3}.\ {#4}. {{\it #5}}, {#7}.\par }
\def\bibb#1#2#3#4#5#6#7{\bibitem{#1} {#2,}\ {#3.}\ {#4.} {#5} {#6} {#7.}}
\def\bibbl#1#2#3#4#5{\bibitem{#1} {#2,}\ {#3.}\ {#4.} {#5.}}
\def\bibC#1#2#3#4#5#6{\bibitem{#1} {#2}\  {#6}.\ :{\ #3},\ {\ #4},\ {{\it #5}}.  \par }

\prep{\end{document}}

\clearpage
\section*{Tables and Figures}
\clearpage

\taba
\clearpage

\tabab
\clearpage

\tabe
\clearpage

\tabf
\clearpage

\tabba
\clearpage

\tabbb
\clearpage

\tabc
\clearpage

\tabd
\clearpage

\figa
\clearpage

\figb
\clearpage

\figc
\clearpage

\figca
\clearpage

\figd
\clearpage

\fige
\clearpage

\figf
\clearpage

\figg
\clearpage

\end{document}